\documentclass[12pt,journal, onecolumn, draftcls]{IEEEtran}
\usepackage{amsmath}
\usepackage{graphicx}
\usepackage{epsfig}
\usepackage{psfrag}
\usepackage{array}
\usepackage{amssymb}
\usepackage{color}

\begin{document}
\title{Efficient Transmit Beamspace Design for Search-free Based DOA Estimation in\\ MIMO Radar}
\author{Arash~Khabbazibasmenj, Aboulnasr Hassanien, Sergiy~A.~Vorobyov, and Matthew~W.~Morency

\thanks{The authors are with the Department of
Electrical and Computer Engineering, University of Alberta,
Edmonton, Alberta, T6G~2V4, Canada (emails: khabbazi@ualberta.ca;
hassanie@ualberta.ca; svorobyo@ualberta.ca; morency@ualberta.ca).
S.~A.~Vorobyov is on leave and currently with the Department of
Signal Processing and Acoustics, Aalto University, Finland.} }

\maketitle

%Low Complexity Direction Finding in Colocated MIMO Radar
%with Arbitrary Receive Array via Transmit Beamspace Design

\begin{abstract}
In this paper, we address the problem of transmit beamspace design
for multiple-input multiple-output (MIMO) radar with colocated
antennas in application to direction-of-arrival (DOA) estimation.
A new method for designing the transmit beamspace matrix that
enables the use of search-free DOA estimation techniques at the
receiver is introduced. The essence of the proposed method is to
design the transmit beamspace matrix based on minimizing the
difference between a desired transmit beampattern and the actual one
under the constraint of uniform power distribution across the
transmit array elements. The desired transmit beampattern can be of
arbitrary shape and is allowed to consist of one or more spatial
sectors. The number of transmit waveforms is even but otherwise arbitrary.
To allow for simple search-free DOA
estimation algorithms at the receive array, the rotational
invariance property is established at the transmit array by
imposing a specific structure on the beamspace matrix. Semidefinite relaxation is used to transform the
proposed formulation into a convex problem that can be solved
efficiently. We also propose a spatial-division based design (SDD) by dividing the
spatial domain into several subsectors and assigning a subset of the
transmit beams to each subsector. The transmit beams associated with each subsector are designed separately. Simulation
results demonstrate the improvement in the DOA estimation
performance offered by using the proposed joint and SDD transmit
beamspace design methods as compared to the traditional MIMO radar
technique.
\end{abstract}


\section{Introduction}
In array processing applications, the direction-of-arrival (DOA)
parameter estimation problem is the most fundamental one \cite{Viberg}. Many
DOA estimation techniques have been developed for the classical
array processing single-input multiple-output (SIMO) setup
\cite{Viberg}, \cite{VanTrees}. The development of a novel array
processing configuration that is best known as multiple-input
multiple-output (MIMO) radar \cite{An_idea}, \cite{Jli} has opened
new opportunities in parameter estimation. Many works have recently
been reported in the literature showing the benefits of applying the
MIMO radar concept using widely separated antennas
\cite{Radar_separated_review}--\cite{MIMORadarSensitivity_Nehorai}
as well as using colocated transmit and receive antennas
\cite{Radar_colocated_review}--\cite{Abeysekera2013}. We focus on
the latter case in this paper.

In MIMO radar with colocated antennas, a virtual array with a larger
number of virtual antenna elements can be formed and used for
improved DOA estimation performance as compared to the performance of SIMO radar
\cite{Duofang}, \cite{NionNikos} for relatively high signal-to-noise
ratios (SNRs), i.e., when the benefits of increased virtual aperture
start to show up. The SNR gain for the traditional MIMO radar (with the
number of waveforms being the same as the number of transmit
 antenna elements), however, decreases as compared to the phased-array radar where the transmit
array radiates a single waveform coherently from all antenna
elements \cite{PhasedMIMOradar}, \cite{EUSIPCO}. A trade-off between
the phased-array and the traditional MIMO radar can be achieved
\cite{PhasedMIMOradar}, \cite{SubarrayedTxBeamformingUK},
\cite{HybridMIMOphased} which gives the best of both configurations,
i.e., the increased number of virtual antenna elements due to the
use of waveform diversity together with SNR gain due to subaperture
based coherent transmission.

Several transmit beamforming techniques have been developed in the
literature to achieve transmit coherent gain in MIMO radar under the
assumption that the general angular locations of the targets are
known a priori to be located within a certain spatial sector. The
increased number of degrees of freedom for MIMO radar, due to the use
of multiple waveforms, is used for the purpose of synthesizing a
desired transmit beampattern based on optimizing the correlation
matrix of the transmitted waveforms \cite{Jli}, \cite{Fuhrmann},
\cite{koivunen}. To apply the designs obtained using the
aforementioned methods, the actual waveforms still have to be found
which can be a difficult and computationally demanding problem
\cite{Stoica}.

One of the major motivations for designing transmit beampattern is
realizing the possibility of achieving SNR gain together with
increased aperture for improved DOA estimation in a wide range of
SNRs \cite{TransmitEnergyFocusing}, \cite{NasrSergCAMSAP}. In
particular, it has been shown in  \cite{TransmitEnergyFocusing} that
the performance of a MIMO radar system with a number of waveforms
less than the number of transmit antennas and with transmit beamspace design capability is better than the performance of a MIMO
radar system with full waveform diversity and no transmit
beamforming gain. Remarkably, using MIMO radar with proper transmit
beamspace design, it is possible to guarantee the
satisfaction of such desired property for DOA estimation as the
rotational invariance property (RIP) at the receive array
\cite{TransmitEnergyFocusing}. This is somewhat similar in effect to
the property of orthogonal space-time block codes in that the shape
of the transmitted constellation does not change at the receiver
independent of the channel. The latter allows for simple decoder
\cite{Tarokh}. Similarly, here the RIP allows for simple DOA
estimation techniques at the receiver although the RIP is actually
enforced at the transmitter, and the propagation media cannot break
it thanks to the proper design of transmit beamspace. Since the RIP holds at
the receive array independent of the propagation media and receive
antenna array configuration, the receive antenna array can be any
arbitrary array. However, the methods developed in
\cite{TransmitEnergyFocusing} suffer from the
shortcomings that the transmit power distribution across the array
elements is not uniform and the achieved phase rotations come with
variations in the magnitude of different transmit beams that
affects the performance of DOA estimation at the receiver.

In this paper, we consider the problem of transmit beamspace design
for DOA estimation in MIMO radar with colocated antennas. We propose
a new method for designing the transmit beamspace that enables the
use of search-free DOA estimation techniques at the
receive antenna array.\footnote{An early and very preliminary exposition of this work has been
presented in parts in \cite{ArashSergiy} and
\cite{SergiyArashMatt}.} The essence of the proposed method is to
design the transmit beamspace matrix based on minimizing the
difference between a desired transmit beampattern and the actual one
while enforcing  the uniform power distribution constraint across
the transmit array antenna elements. The desired transmit beampattern can be
of arbitrary shape and is allowed to consist of one or more spatial
sectors. The case of even but otherwise arbitrary number of transmit
waveforms is considered. To allow for simple search-free DOA
estimation algorithms at the receiver, the RIP is established at the
transmit antenna array by imposing a specific structure on the transmit
beamspace matrix. The proposed structure is based on designing the
transmit beams in pairs where the transmit weight vector associated
with a certain transmit beam is the conjugate flipped version of the
weight vector associated with another beam, i.e., one transmit
weight vector is designed for each pair of transmit beams. All pairs
are designed jointly while satisfying the requirement that the two
transmit beams associated with each pair enjoy rotational invariance
with respect to each other. Semidefinite programming (SDP) relaxation is used to
transform the proposed formulation into a convex problem that can
be solved efficiently  using, for example, interior point methods.
In comparison to our previous method \cite{NasrSergCAMSAP} that achieves phase rotation
between two transmit beams, the proposed method enjoys the following
advantages. (i) It ensures that the magnitude response of the two
transmit beams associated with one pair of transmit beams is exactly the same at all
spatial directions, a property that improves the DOA estimation
performance. (ii) It ensures uniform power distribution across
transmit elements. (iii) It
enables estimating the DOAs via estimating the accumulated phase
rotations over all transmit beams instead of only two beams. (iv) It
only involves optimization over half the entries of the transmit
beamspace matrix which decreases the computational load. We also
propose an alternative formulation based on splitting the overall
transmit beamspace design problem into several smaller problems.
The alternative formulation is referred to as the spatial-division
based design (SDD) which involves dividing the spatial domain into
several subsectors and assigning a subset of the transmit beamspace
pairs to each subsector. The SDD method enables post processing of
data associated with different subsectors independently with
estimation performance comparable to the performance of the joint
transmit beamspace design. Simulation results demonstrate the
improvement in the DOA estimation performance that is achieved
by using the proposed joint transmit beamspace design and SDD
methods as compared to the traditional MIMO radar technique.

The rest of the paper is organized as follows. Section~II introduces
the system model for mono-static MIMO radar system with transmit
beamspace. The problem formulation is developed in Section~III while the transmit beamspace design problem for even but
otherwise arbitrary number of transmit waveforms is developed in
Section~IV. Section~V gives simulation examples for the proposed
DOA estimation techniques and conclusions are drawn in Section~VI.

\section{System Model and Main Idea}

Consider a mono-static MIMO radar system equipped with a uniform linear transmit
array of $M$ colocated antennas with inter-element spacing $d$
measured in wavelength and a receive array of $N$ antennas
configured in a random shape. The transmit and receive arrays are
assumed to be close enough to each other such that the spatial angle
of a target in the far-field remains the same with respect to both
arrays. Let ${\boldsymbol\Phi}(t)=[\phi_1(t),\ldots, \phi_K(t)]^T$ be
the $K\times 1$ vector that contains the complex envelopes of the
waveforms $\phi_k(t),\ k=1,\ldots,K$ which are assumed to be
orthogonal, i.e.,
\begin{equation}\label{eq:orthogonality}
\int_{0}^{\rm T_p}{\phi_i(t)\phi_j^{*}}(t) = \delta(i-j), \qquad
i,j=1,2,\cdots,K
\end{equation}
where $T_p$ is the pulse duration, $(\cdot)^T$ and $(\cdot)^*$ stand for the transpose and the conjugate, respectively, and $\delta (\cdot )$ is the
Kroneker delta. The actual transmitted signals are taken as linear combinations of the orthogonal waveforms. Therefore, the $M \times 1$ vector of
the baseband representation of the transmitted signals can be written as
\cite{TransmitEnergyFocusing}
\begin{equation} \label{TS}
\mathbf{s} (t) = \left[ s_1(t),\ldots, s_M(t)\right]^T
               = \mathbf{W} {\boldsymbol \Phi} (t)
\end{equation}
where $s_i(t)$ is the signal transmitted from antenna $i$ and
\begin{equation}\label{eq:TxWeightMatrix}
\mathbf W  =
         \begin{pmatrix}
            w_{1,1} & w_{2,1} & \cdots & w_{K,1} \\
            w_{1,2} & w_{2,2} & \cdots & w_{K,2} \\
            \vdots  & \vdots  & \ddots & \vdots  \\
            w_{1,M} & w_{2,M} & \cdots & w_{K,M}
         \end{pmatrix}
\end{equation}
is the $M\times K$ transmit beamspace matrix. It is worth noting that
each of the orthogonal waveforms $\phi_k(t),\ k=1,\ldots,K$ is
transmitted over one transmit beam where the $k$th column of the
matrix $\mathbf W$ corresponds to the transmit beamforming weight
vector used to form the $k$th beam.

Let $\mathbf{a} (\theta) \triangleq [1, e^{-j 2 \pi  d \sin (\theta)}, \ldots , e^{-j 2 \pi   d (M-1) \sin (\theta)
}]^T$ be the $M\times 1$ transmit array steering vector. The transmit power distribution pattern can be expressed as \cite{Fuhrmann}
\begin{equation}
\label{pattern} G(\theta)=\frac{1}{4 \pi} \mathbf d^H(\theta)
\mathbf R \mathbf d(\theta),\quad -\pi/2\leq \theta \leq \pi/2
\end{equation}
where $(\cdot)^H$ stands for the conjugate transpose, $\mathbf{d}(\theta) =$ $\mathbf{a}^{*}(\theta)$, and
\begin{equation}\label{eq:cross-correlation}
{\bf R} = \int_0^{\rm T_ p} {\bf s}(t) {\bf s}^H(t) dt
\end{equation}
is the cross-correlation matrix of the transmitted signals \eqref{TS}.
One way to achieve a certain desired transmit beampattern is to
optimize over the cross-correlation matrix ${\bf R}$ such as in
\cite{Fuhrmann}, \cite{koivunen}. In this case, a complementary problem has to be
solved after obtaining ${\bf R}$ in order to find appropriate signal
vector ${\bf s}(t)$ that satisfies \eqref{eq:cross-correlation}.
Solving such a complementary problem is in general difficult and
computationally demanding. However, in this paper, we extend our approach of optimizing
the transmit beampattern via designing the transmit beamspace matrix.
According to this approach, the cross-correlation matrix is expressed as
\begin{equation}\label{eq:cross-correlation1}
\mathbf R = \mathbf W \mathbf W^{H}
\end{equation}
that holds due to the orthogonality of the waveforms (see \eqref{eq:orthogonality} and \eqref{TS}).
Then the transmit beamspace matrix $\mathbf W$ can be designed to achieve the desired
beampattern while satisfying many other requirements mandated by
practical considerations such as equal transmit power distribution
across the transmit array antenna elements, achieving a desired radar ambiguity function,
etc. Moreover, this approach enables enforcing the RIP which
facilitates subsequent processing steps at the receive antenna array, e.g., it
enables applying accurate computationally efficient DOA estimation
using search-free direction finding techniques such as ESPRIT.

The signal measured at the output of the receive array due to echoes
from $L$ narrowband far-field targets can be modeled as
\begin{equation}\label{eq:receivedData}
\mathbf x (t, \tau ) = \sum_{l=1}^L \beta_l(\tau) \left[ \mathbf d^H(\theta_l)\mathbf W
{\boldsymbol \Phi}(t)\right] \mathbf b ( \theta_l )  + \mathbf z ( t, \tau )
\end{equation}
where $t$ is the time index within the radar pulse, $\tau$ is the
slow time index , i.e., the pulse number, $\beta_l(\tau)$ is the
reflection coefficient of the target located at the unknown spatial
angle $\theta_l$, $\mathbf b ( \theta_l )$ is the receive array
steering vector, and $\mathbf z ( t, \tau)$ is the $N \times 1$
vector of zero-mean white Gaussian noise with variance $\sigma_z^2$.
In \eqref{eq:receivedData}, the target reflection coefficients
$\beta_l(\tau),\ l=1,\ldots,L$ are assumed to obey the Swerling~II
model, i.e, they remain constant during the duration of one radar
pulse but change from pulse to pulse. Moreover, they are assumed to
be drawn from a normal distribution with zero mean and variance
$\sigma_\beta^2$.

By matched filtering $\mathbf x(t, \tau )$ to each
of the orthogonal basis waveforms $\phi_k(t), k=1,\ldots,K$, the
$N\times 1$ virtual data vectors can be obtained
as\footnote{Practically, this matched filtering step is performed
for each Doppler-range bin, i.e., the received data $\mathbf x(t,
\tau )$ is matched filtered to a time-delayed Doppler shifted
version of the waveforms $\phi_k(t), k=1,\ldots,K$.}
\begin{eqnarray}\label{eq:virtualData}
\!\!\!\mathbf y_{k} (\tau)\!&\!\!\!=\!\!\!&\!\int_{\rm T_p}
\mathbf x(t,\tau) \phi_k^* (t) dt\nonumber\\
\!&\!\!\!=\!\!\!&\! \sum_{l=1}^L \beta_l(\tau) \left({\mathbf d}^H(\theta_l){\mathbf w}_k\right)
\mathbf b(\theta_l) + {\mathbf z}_k(\tau)
\end{eqnarray}
where ${\mathbf w}_k$ is the $k$th column of the transmit beamspace matrix
$\mathbf W$  and ${\mathbf z}_k(\tau) \triangleq \int_{\rm T_{\rm p}} \mathbf z(t,\tau) \phi_k^* (t) dt$ is the
$N \times 1$  noise term whose covariance is $ \sigma_z^2 \mathbf
I_{N}$.

Let $\breve{\mathbf y}_{l,k}(\tau)$ be the noise free component of
the virtual data vector \eqref{eq:virtualData} associated with the
$l$th target, i.e., $\breve{\mathbf y}_{l,k}(\tau) = \beta_l(\tau)
\left({\mathbf d}^H(\theta_l){\mathbf w}_k\right) \mathbf
b(\theta_l)$. Then, one can easily observe that the $k$th  and the
$k^{\prime}$th components associated with the $l$th target are
related to each other through the following relationship

\begin{eqnarray} \label{eq:virtualData1}
\breve{\mathbf y}_{l,k^{\prime}} (\tau) \!&\!\!\!=\!\!\!&\! \beta_l(\tau) \left({\mathbf d}^H(\theta_l){\mathbf w}_{k^{\prime}}\right) \mathbf b(\theta_l) \nonumber\\
\!&\!\!\!=\!\!\!&\! \frac{{\mathbf d}^H(\theta_l){\mathbf w}_{k^{\prime}}}{{\mathbf d}^H(\theta_l){\mathbf w}_{k}} \cdot \breve{\mathbf y}_{l,k} (\tau)\nonumber\\
\!&\!\!\!=\!\!\!&\! e^{j \left( \psi_{k^{\prime}}(\theta_l) - \psi_k(\theta_l) \right)} \frac{\left| {\mathbf d}^H(\theta_l){\mathbf w}_{k^{\prime}} \right|}{ \left| {\mathbf d}^H(\theta_l){\mathbf w}_{k} \right| } \cdot \breve{\mathbf y}_{l,k} (\tau)
\end{eqnarray}
where $\psi_k(\theta)$ is the phase of the inner product ${{\mathbf d}^H(\theta)\mathbf
w}_k$. The expression~\eqref{eq:virtualData1} means that
the signal component  $y_{k}(\tau)$ corresponding to a given target
is the same as the signal component $\mathbf y_{k^{\prime}}$
corresponding to the same target up to a phase rotation and a gain
factor.

The RIP can be enforced by imposing the constraint
$|{\mathbf d}^H(\theta){\mathbf w}_k| = |{\mathbf d}^H(\theta){\mathbf
w}_{k^{\prime}}|$ while designing the transmit
beamspace matrix $\mathbf W$. The main advantage of enforcing the
RIP is that it allows us to estimate DOAs via estimating the phase
rotation associated with the $k$th and $k^{\prime}$th pair of the
virtual data vectors using search-free techniques, e.g., ESPRIT.
Moreover, if the number of transmit waveforms is more than two, the
DOA estimation can be carried out via estimating the phase
difference
\begin{equation}\label{eq:PhaseAccumulation}
\angle \sum_{i=1}^{K/2} {\mathbf d}^H( \theta_l)\mathbf w_{i} - \!\angle\!\!\!\sum_{i=K/2+1}^{K}\!\!\! {\mathbf
d}^H(\theta_l)\mathbf w_{i}
\end{equation}
and comparing it to a precalculated phase profile for the given
spatial sector in which we have concentrated power from the transmit
antenna array. However, in the latter case, precautions should be taken to
assure the coherent accumulation of the $K/2$ components in
\eqref{eq:PhaseAccumulation}, i.e., to avoid gain loss as will be
shown later in the paper.

\section{Problem formulation}
The main goal is to design a transmit beamspace matrix $\mathbf
W$ which achieves a spatial beampattern that is as close as
possible to a certain desired one. Substituting
\eqref{eq:cross-correlation1} in (\ref{pattern}), the spatial
beampattern can be rewritten as
\begin{eqnarray}\label{eq:beampattern1}
G(\theta) &\!\!\!=\!\!\!& \frac{1}{4 \pi} \mathbf d^H(\theta) \mathbf W \mathbf
W^H \mathbf d(\theta)\nonumber\\
&\!\!\!=\!\!\!& \frac{1}{4 \pi} \sum_{i=1}^{K} \mathbf
w_i^H \mathbf d(\theta) \mathbf d^H (\theta) \mathbf w_i.
\end{eqnarray}
Therefore, we design the transmit beamspace matrix $\mathbf W$ based on
minimizing the difference between the desired beampattern and the
actual beampattern given by \eqref{eq:beampattern1}. Using the minmax criterion, the transmit
beamspace matrix design problem can be formulated as
\begin{eqnarray}
\!\!\!\!\!&\!\!\!\!\!\!&\!\!\!\min_{\mathbf W} \max_{\theta} \left| G_{\rm d}(\theta) - \frac{1}{4 \pi}
 \sum_{i=1}^{K} \mathbf w_i^H \mathbf d (\theta) \mathbf d^H (\theta)
 \mathbf w_i \right| \label{objective}  \\
\!\!\!\!\!&\!\!\!\!\!\!&\!\!\! {\rm s.t.}  \ \ \   \sum _{i=1}^{K} |\mathbf w_{i}(j)|^2=\frac{P_{\rm t}}{M}, \quad
  j=1,\cdots,M  \label{antenna}% \\
\end{eqnarray}
where $ G_{\rm d}(\theta), \theta \in [- \pi /2 , \pi /2]$ is the desired beampattern and $P_{\rm t}$ is the total transmit power. The $M$ constraints enforced in (\ref{antenna}) are used to ensure that individual antennas transmit equal powers given by $P_{\rm t}/M$. It is equivalent to having the  norms of the rows of ${\mathbf W}$ to be equal to $P_{\rm t}/M$. The uniform power distribution across the array antenna elements given by (\ref{antenna}) is necessary from a practical point of view. In practice, each antenna in the transmit array typically uses the same power amplifier, and thus has the same dynamic power range. If the power used by different antenna elements is allowed to vary widely, this can severely degrade the performance of the system due to the nonlinear characteristics of the power amplifier.

Another goal that we wish to achieve is to enforce the RIP to enable
for search-free DOA estimation. Enforcing the RIP between the $k$th
and $(K/2+k)$th transmit beams is equivalent to ensuring that the following
relationship holds
\begin{equation}
\left|{\mathbf w}_k^H{\mathbf d}(\theta)\right| = \left|{\mathbf
w}_{\frac{K}{2}+k}^H{\mathbf d}(\theta)\right|, \quad \theta \in
[-\pi/2,\pi/2]. \label{TheRIP}
\end{equation}
Ensuring \eqref{TheRIP}, the optimization problem
\eqref{objective}--\eqref{antenna} can be reformulated as
\begin{eqnarray}
\!\!\!\!\!&\!\!\!\!\!\!&\!\!\!\min_{\mathbf W} \max_{\theta} \left| G_{\rm d}(\theta) - \frac{1}{4 \pi}
 \sum_{i=1}^{K} \mathbf w_i^H \mathbf d (\theta) \mathbf d^H (\theta)
 \mathbf w_i \right| \label{objective1}  \\
\!\!\!\!\!&\!\!\!\!\!\!&\!\!\! {\rm s.t.}  \ \ \   \sum _{i=1}^{K} |\mathbf w_{i}(j)|^2=\frac{P_{\rm t}}{M}, \quad
  j=1,\cdots,M  \label{antenna1} \\
\!\!\!\!\!&\!\!\!\!\!\!&\!\!\!\qquad\ \ \ \left|{\mathbf w}_k^H{\mathbf d}(\theta)\right| = \left|{\mathbf
w}_{\frac{K}{2}+k}^H{\mathbf d}(\theta)\right|, \label{TheRIP1} \\
\!\!\!\!\!&\!\!\!\!\!\!&\!\!\!\qquad\ \ \  \theta \in [-\pi/2,\pi/2],\quad k=1,\ldots,\frac{K}{2}\nonumber.
\end{eqnarray}
It is worth noting that the constraints \eqref{antenna1} as well as
the constraints \eqref{TheRIP1} correspond to non-convex sets and,
therefore, the optimization problem
\eqref{objective1}--\eqref{TheRIP1} is a non-convex problem which is
difficult to solve in a computationally efficient manner. Moreover,
the fact that \eqref{TheRIP1} should be enforced for every direction
$\theta \in [-\pi/2,\pi/2]$, i.e., the number of equations in
\eqref{TheRIP1} is significantly larger than the number of the
variables, makes it impossible to satisfy \eqref{TheRIP1} unless a
specific structure on the transmit beamspace matrix $\mathbf W$ is
imposed.

In the following section we propose a specific structure to $\mathbf
W$ to overcome the difficulties caused by  \eqref{TheRIP1} and show
how to use SDP relaxation to overcome the difficulties
caused by the non-convexity of \eqref{objective1}--\eqref{TheRIP1}.

\section{Transmit beamspace design}

\subsection{Two Transmit Waveforms}
\label{BDes2Wave} We first consider a special, but practically
important case of two orthonormal waveforms. Thus, the dimension of
$\mathbf W$ is{ $ M \times 2$}. Then under the aforementioned
assumption of ULA at the MIMO radar transmitter, the RIP can be
satisfied by choosing the transmit beamspace matrix to take the form
\begin{equation}
\mathbf W = \left[\mathbf w, \, \tilde{\mathbf w}^*\right]\label{eq:SpecialStructure}
\end{equation}
where $\tilde{\mathbf w}$ is the flipped version of vector $\mathbf w$,
i.e., $\tilde{\mathbf w} (i) = \mathbf w (M-i+1)$, $i = 1, \ldots,
M$. Indeed, in this case, $| {\mathbf d}^H (\theta) {\mathbf w} | = |
{\mathbf d}^H(\theta) {\tilde{\mathbf w}^*} |$ and the RIP is clearly
satisfied.

To prove that the specific structure \eqref{eq:SpecialStructure}
achieves the RIP, let us represent the vector $\mathbf w$ as a vector of complex numbers
\begin{equation}
\mathbf w = [z_1\ z_2 \ldots z_M]^T
\end{equation}
where $z_m,\ m=1,\ldots,M$ are complex numbers. Then the
flipped-conjugate version of $\mathbf w$  has the structure
${\tilde{\mathbf w}}^{*} = [z_M^{*}\ z_{M-1}^{*} \ldots z_1^{*}]^T$.
Examining the inner products ${\mathbf d}^H(\theta) {\bf w}$ and
${\mathbf d}^H(\theta) { \tilde{\mathbf w}}^{*}$ we see that the
first inner product produces the sum
\begin{eqnarray}
{\mathbf d}^H(\theta) {\bf w} &\!\!\!=\!\!\!& z_1 + z_2 e^{j 2 \pi  d \sin(\theta)} + \nonumber\\
&\!\!\! \!\!\!& \ldots + z_M e^{j2 \pi  d \sin (\theta)(M-1)}
\end{eqnarray}
and the second produces the sum
\begin{eqnarray}
{\mathbf d}^H(\theta) { \tilde{\mathbf w}_i}^{*} &\!\!\!=\!\!\!& z_M^{*} +
z_{M-1}^{*}\cdot e^{j 2 \pi  d \sin(\theta)} + \nonumber\\
&\!\!\! \!\!\!& \ldots + z_1^{*}\cdot e^{j 2 \pi d \sin
(\theta)(M-1)}.\label{eq:SecondProduct}
\end{eqnarray}
Factoring out the term $e^{-j 2 \pi  d \sin (\theta)(M-1)}$ from \eqref{eq:SecondProduct} and conjugating, we can see that
the sums are identical in magnitude and indeed are the
same up to a phase rotation $\psi$. This relationship is precisely
the RIP, and it is enforced at the transmit antenna array by the structure imposed on the transmit beamspace matrix $\mathbf
W$.

Substituting \eqref{eq:SpecialStructure} in \eqref{objective1}--\eqref{TheRIP1}, the optimization problem can be reformulated for the case of two transmit waveforms as follows
\begin{eqnarray}
&&\min_{\mathbf w} \max_{ \theta} \left| G_{\rm d} (\theta) -
 \|  [ \mathbf w \, \tilde{\mathbf w}^* ]^H \mathbf d (\theta) \|^2  \ \right| \label{objective2}\\
&& {\rm s.t.}  \, | \mathbf w (i) |^2 + | \tilde{\mathbf w} (i)
|^2 = \frac{P_{\rm t}}{M}, \ \ \ i=1,\ldots,M.  \label{ME_PC}
\end{eqnarray}
It is worth noting that the constraints \eqref{TheRIP1} are not
shown in the optimization problem \eqref{objective2}--\eqref{ME_PC}
because they are inherently enforced due to the use of the specific
structure of $\mathbf W$ given in \eqref{eq:SpecialStructure}.

Introducing the auxiliary variable $\delta$, the optimization problem
\eqref{objective2}--\eqref{ME_PC} can be equivalently rewritten as
\begin{eqnarray}
\!\!\!\!\!\!\!\!\!\!\!\!\!\!\!&&\min_{\mathbf w, \delta} \delta \nonumber\\
\!\!\!\!\!\!\!\!\!\!\!\!\!\!\!&& {\rm s.t.} \ \ \
\frac{G_{\rm d}(\theta_q)}{2} \!-\!|  \mathbf w^H \mathbf d
(\theta_q)|^2  \leq
\delta, \, q = 1, \ldots, Q \nonumber \\
\!\!\!\!\!\!\!\!\!\!\!\!\!\!\!&& \qquad \frac{G_{\rm d}(\theta_q)}{2}
\!-\! |  \mathbf w^H \mathbf d (\theta_q) |^2 \geq -\delta, \,
q = 1, \ldots, Q \nonumber \\
\!\!\!\!\!\!\!\!\!\!\!\!\!\!\!&& \qquad | \mathbf w (i) |^2 \!+\!
| \mathbf w (M\!-i\!+\!1) |^2 \!= \!\frac{P_{\rm t}}{M}, \, i = 1, \ldots
\!, \frac{M}{2}. \label{conss3}
\end{eqnarray}
where $\theta_q \in [-\pi/2,\pi/2],\ q=1,\ldots,Q$ is a continuum of
directions that are properly chosen (uniform or nonuniform) to
approximate the spatial domain $[-\pi/2,\pi/2]$. It is worth noting
that the optimization problem \eqref{conss3} has significantly
larger number of degrees of freedom than the beamforming problem for
the phased-array case where the magnitudes of $\mathbf w (i)$, $i =
1, \ldots, M$ are fixed.

The problem \eqref{conss3} belongs to the
class of non-convex quadratically-constrained quadratic programming
(QCQP) problems which are in general NP-hard. However, a well
developed SDP relaxation technique can be used to solve it \cite{SDP2}--\cite{SDP3}. Indeed, using the
facts that $|  \mathbf w^H \mathbf d (\theta_q)|^2 = {\rm tr} (
\mathbf d (\theta_q ) \mathbf d^H (\theta_q) \mathbf w \mathbf w^H)$
and $|\mathbf w (i) |^2 + | \mathbf w (M-i+1) |^2 = {\rm tr}(
\mathbf w \mathbf w^H \mathbf A_i), i=1,\ldots,M/2$, where ${\rm
tr}(\cdot)$ stands for the trace and $\mathbf A_i$ is an $M\times M$
matrix such that $\mathbf A_i (i,i) = \mathbf A_i (M-(i-1), M-(i-1))
= 1$ and the rest of the elements are equal to zero, the problem
\eqref{conss3} can be cast as
\begin{eqnarray}
\!\!\!\!\!\!\!\!\!\!\!\!\!\!\!&&\min_{\mathbf w, \delta} \delta
\nonumber \\
\!\!\!\!\!\!\!\!\!\!\!\!\!\!\!&& {\rm s.t.} \ \ \ \frac{G_{\rm d} (
\theta_q )}{2} \!-\! {\rm tr} ( \mathbf d (\theta_q) \mathbf d^H
(\theta_q) \mathbf w \mathbf w^H ) \!\leq\! \delta, \, q = 1,
\ldots, Q \nonumber \\
\!\!\!\!\!\!\!\!\!\!\!\!\!\!\!&& \qquad \frac{G_{\rm d}(\theta_q)}{2}
\!-\! {\rm tr} (\mathbf d (\theta_q) \mathbf d^H (\theta_q) \mathbf
w \mathbf w^H) \!\geq\!\! -\delta, \, q = 1,
\ldots, Q \nonumber \\
\!\!\!\!\!\!\!\!\!\!\!\!\!\!\!&& \qquad {\rm tr} ( \mathbf w \mathbf
w^H \mathbf A_i ) = \frac{P_{\rm t}}{M}, \, i = 1, \ldots, \frac{M}{2}.
\label{newcon}
\end{eqnarray}

Introducing the new variable $ \mathbf X \triangleq \mathbf w
\mathbf w^H $, the problem \eqref{newcon} can be equivalently
written as
\begin{eqnarray}
\!\!\!\!\!\!\!\!\!\!\!\!\!\!\!&& \min_{\mathbf X, \delta} \delta
\nonumber \\
\!\!\!\!\!\!\!\!\!\!\!\!\!\!\!&& {\rm s.t.} \ \ \ \frac{G_{\rm d} (
\theta_q)}{2} \!-\! {\rm tr} ( \mathbf d(\theta_q) \mathbf d^H (
\theta_q ) \mathbf X) \!\leq\! \delta, \, q = 1, \ldots, Q \nonumber \\
\!\!\!\!\!\!\!\!\!\!\!\!\!\!\!&& \qquad \frac{G_{\rm d}(\theta_q)}{2}
\!-\! {\rm tr} ( \mathbf d(\theta_q) \mathbf d^H (\theta_q)
\mathbf X) \!\geq\! -\delta, \, q = 1, \ldots, Q \nonumber \\
\!\!\!\!\!\!\!\!\!\!\!\!\!\!\!&& \qquad {\rm tr} ( \mathbf X \mathbf
A_i)= \frac{P_{\rm t}}{M}, \, i = 1, \ldots, \frac{M}{2}; \; {\rm rank}( \mathbf
X) =1 \label{rank}
\end{eqnarray}
where $\mathbf X$ is the Hermitian matrix and ${\rm rank}(\cdot)$
denotes the rank of a matrix. Note that the last two constraints
in \eqref{rank} imply that the matrix $\mathbf X$ is positive
semidefinite. The problem \eqref{rank} is non-convex with respect
to $\mathbf X$ because the last constraint is not convex. However,
by means of the SDP relaxation technique, this constraint can be
replaced by another constraint, that is, $\mathbf X \succeq
{\bf 0}$. The resulting problem is the relaxed version of \eqref{rank}
and it is a convex SDP problem which can be efficiently solved
using, for example, interior point methods. When the relaxed
problem is solved, extraction of the solution of the original
problem is typically done via the so-called \textit{randomization}
techniques \cite{SDP2}.

Let $\mathbf X_{\rm opt}$ denote the optimal solution of the relaxed
problem. If the rank of $\mathbf X_{\rm opt}$ is one, the optimal
solution of the original problem \eqref{conss3} can be obtained by
simply finding the principal eigenvector of $\mathbf X_{\rm opt}$.
However, if the rank of the matrix $\mathbf X_{\rm opt}$ is higher
than one, the randomization approach can be used. Various
randomization techniques have been developed and are generally
based on generating a set of candidate vectors and then choosing
the candidate which gives the minimum of the objective function of
the original problem. Our randomization procedure can be described
as follows. Let $\mathbf X_{\rm opt} = \mathbf U \mathbf \Sigma
\mathbf U^H$ denote the eigen-decomposition of $\mathbf X_{\rm opt}$.
The candidate vector $k$ can be chosen as $\mathbf w_{{\rm can},k}=
\mathbf U \mathbf \Sigma^{1/2} \mathbf v_k$ where $\mathbf v_k$ is
random vector whose elements are random variables uniformly
distributed on the unit circle in the complex plane. Candidate
vectors are not always feasible and should be mapped to a nearby
feasible point. This mapping is problem dependent \cite{SDP3}. In
our case, if the condition $| \mathbf w_{{\rm can},k} (i) |^2 + |
\mathbf w_{{\rm can},k} (M-i+1) |^2 = P_{\rm t}/M$ does not hold, we can map
this vector to a nearby feasible point by scaling $\mathbf
w_{{\rm can},k} (i)$ and $\mathbf w_{{\rm can},k} (M-i+1)$ to satisfy this
constraint. Among the candidate vectors we then choose the one
which gives the minimum objective function, i.e., the one with
minimum $\max_{\theta_q} \left|{G_{\rm d}(\theta_q)}/{2}  - |\mathbf
w_{{\rm can},k}^H \mathbf d(\theta_q) |^2\right|$.

\subsection{Even Number of Transmit Waveforms}
Let us consider now the {$ M \times K$} transmit beamspace matrix
$\mathbf W=[\mathbf w_1,  \mathbf w_2,  \cdots, \mathbf w_K]$ where
$K \leq M$ and $K$ is an even number. For convenience, the virtual
received signal vector matched to the basis waveform $\phi_k(t)$ is rewritten as
\begin{eqnarray}
\!\!\!\mathbf y_{k} ( \tau)\!&\!\!\!=\!\!\!&\!\int_{\rm T_p}
\mathbf x(t, \tau) \phi_k^* (t) dt\nonumber\\
\!&\!\!\!=\!\!\!&\! \sum_{l=1}^L \beta_l(\tau) e^{j\psi_k
(\theta_l)}\left|{\mathbf d}^H(\theta_l){\mathbf w}_k\right| \mathbf
b(\theta_l) + {\mathbf z}_k( \tau) . \label{virtualrepeated}
\end{eqnarray}
From \eqref{virtualrepeated}, it can be seen that the RIP between
$\mathbf y_{k}$ and $\mathbf y_{k^{\prime}}, k\neq k^{\prime}$ holds if
\begin{equation}
\left| {\mathbf d}^H(\theta){\mathbf w}_k \right| = \left| {\mathbf d}^H(\theta){\mathbf
w}_{k^{\prime}} \right|, \qquad \theta \in
[-\pi/2,\pi/2] . \label{2Wave}
\end{equation}

In the previous subsection, we saw that by considering the following
specific structure  $[\mathbf w \ \ \tilde{\mathbf w}^*]$ for the
transmit beamspace matrix with only two waveforms, the RIP is guaranteed at the receive antenna array.
In this part, we obtain the RIP for the more general case of more than two
waveforms. It provides more degrees of freedom for obtaining
a better performance. For this goal, we first show that if for some
$k^\prime$ the following relation holds
\begin{equation}
\left| \sum_{i=1}^{k^\prime} {\mathbf d}^H(\theta) \mathbf w_i \right| = \left| \sum_{i=k^\prime+1}^{K} {\mathbf d}^H(\theta) \mathbf w_i \right|, \ \ \forall \theta \in [-\pi/2,\pi/2]
\label{BasePro}
\end{equation}
then the two new sets of vectors defined as the summation of the
first $k^{\prime}$ data vectors $\mathbf y_{i} ( \tau)$, $\
i=1,\cdots,k^{\prime}$ and the last
$K-k^{\prime}$ data vectors $\mathbf y_{i} ( \tau), \
i=k^{\prime}+1,\cdots,K$  will satisfy the RIP. More specifically,
by defining the following vectors
\begin{eqnarray}
{\mathbf g}_1(\tau) \!\!\!&\!\!\triangleq\!\!&\!\!\!
\sum_{i=1}^{k^{\prime}} \mathbf y_{i} ( \tau) \nonumber \\
\!\!\!\!&\!\!\!=\!\!\!& \!\!\!\!\!\! \sum_{l=1}^L\!\! \beta_l( \tau) \! \left (
\sum_{i=1}^{k^{\prime}} {\mathbf d}^H(\theta_l){\mathbf w}_i \right)
\mathbf b (\theta_l) \!+\! \sum_{i=1}^{k^{\prime}}
{\mathbf z}_i( \tau) \\
{\mathbf g}_2(\tau) \!\!\!&\!\!\!\triangleq\!\!\!&\!\!\!
\sum_{i=k^{\prime}+1}^{K} \mathbf y_{i} ( \tau) \nonumber \\
\!\!\!\!&=& \!\!\!\!\!\!  \sum_{l=1}^L \beta_l( \tau) \left (
\sum_{i\!=\!k^{\prime}\!+\!1}^{K} \!\!{\mathbf
d}^H(\theta_l) {\mathbf w}_i\!\! \right)\!\mathbf b(\theta_l)\! + \!\!\!\!\!
\sum_{i=k^{\prime}\!+\!1}^{K}\!\!\!{\mathbf z}_i( \tau)
\end{eqnarray}
the corresponding signal component of target $l$ in the vector
${\mathbf g}_1(\tau)$ has the same magnitude as in the vector ${\mathbf g}_2(\tau)$
if the equation \eqref{BasePro} holds. In this case, the only
difference between the signal components of the target $l$ in the
vectors ${\mathbf g}_1(\tau)$ and ${\mathbf g}_2(\tau)$ is the phase which can be
used for DOA estimation. Based on this fact, for ensuring the
RIP between the vectors ${\mathbf g}_1(\tau)$ and ${\mathbf g}_2(\tau)$, equation \eqref{BasePro} needs to be satisfied for every angle
$\theta \in [-\pi/2,\pi/2]$. By noting that the equation $| {\mathbf d}^H(\theta){\mathbf w} | = | {\mathbf
d}^H(\theta) \tilde{\mathbf w}^* |$ holds for any arbitrary $\theta$, it can be shown that
the equation \eqref{BasePro} holds for any arbitrary $\theta$ only
if the following structure on the matrix $\mathbf W$ is imposed:
\begin{itemize}
\item $K$ is an even number,
\item $k^{\prime}$ equals to $K/2$,
\item $\mathbf w_i ={ \tilde{\mathbf w}_{k^{\prime}+i}}^{*}, \ \ \
i=1,\cdots,K/2$.
\end{itemize}
More specifically, if the transmit beamspace matrix has the
following structure
\begin{equation}
\label{GeneralFormStructure} \mathbf W = [\mathbf w_1,\cdots,\mathbf
w_{K/2},\tilde{\mathbf w}_1^*,\cdots,\tilde{\mathbf w}_{K/2}^*]
\end{equation}
then the signal component of ${\mathbf g}_1(\tau)$ associated with the $l$th
target is the same as the corresponding signal component of ${\mathbf g}_2(\tau)$ up to phase rotation of
\begin{equation}
\angle \sum_{i=1}^{K/2} {\mathbf d}^H(\theta_l)\mathbf w_i -\!\angle\!\!\!\!
\sum_{i=K/2+1}^{K}\!\!\!\! {\mathbf d}^H(\theta_l)\mathbf w_i
\end{equation}
which can be used as a look-up table for finding DOA of a target. By
considering the aforementioned structure for the transmit beamspace matrix
$\mathbf W$, it is guaranteed that the RIP is satisfied and other additional design requirements can
be satisfied through the proper design of $\mathbf
w_1,\cdots,\mathbf w_{K/2}$.

Substituting \eqref{GeneralFormStructure} in \eqref{TheRIP1}, the
optimization problem of transmit beamspace  matrix
design can be reformulated as
\begin{eqnarray}
\!\!\!&\!\!\!\min\limits_{\mathbf w_k} \max\limits_{ \theta_q}\!\!\!&\!\!\! \left| G_{\rm d}
(\theta_q) - \sum_{k=1}^{K/2} \|  [ \mathbf w_k \, \tilde{\mathbf
w_k}^* ]^H
\mathbf d (\theta_q) \|^2 \right|\label{E_PCM10problem} \\
\!\!\!&\!\!\! {\rm s.t.} \!\!\!&\!\!\!  \, \sum_{k=1}^{K/2} | \mathbf w_k (i) |^2 + |
\tilde{\mathbf w}_k (i) |^2 \!=\! \frac{P_{\rm t}}{M}, \quad i=1,\ldots,M.
\nonumber
\end{eqnarray}
For the case when the number of transmit antennas is even\footnote{The case when the number of transmit antennas is odd can be carried out in a straightforward manner.} and using
the facts that
\begin{equation}
\label{two2one} \| [ \mathbf w_k \, \tilde{\mathbf w}_k^* ]^H
\mathbf d (\theta_q) \|^2=2|\mathbf w_k ^H  \mathbf d (\theta_q) |^2
\end{equation}
\begin{equation}
 |  \mathbf w_k^H \mathbf d (\theta_q)|^2 = {\rm tr} (
\mathbf d (\theta_q ) \mathbf d^H (\theta_q) \mathbf w_k \mathbf
w_k^H)
\end{equation}
\begin{eqnarray}
|\mathbf w_k (i) |^2 + | \mathbf w_k (M-i+1) |^2 \!\!\!&\!\!\!=\!\!\!&\!\!\!   {\rm tr}(
\mathbf w_k  \mathbf w_k^H \mathbf A_i),  \nonumber\\ && \quad
i=1,\ldots,M/2
\end{eqnarray}
the problem \eqref{E_PCM10problem} can be recast as
\begin{eqnarray}
 &\min\limits_{\mathbf w_k}
\max\limits_{\theta_q}& \left| G_{\rm d} (\theta_q)/2 - \sum_{k=1}^{K/2}
\left|{\mathbf d}^H (\theta_q) \mathbf w_k  \right|^2
\right| \nonumber \\
&{\rm s.t.}& \sum_{k=1}^{K/2} {\rm tr} ( \mathbf w_k \mathbf w_k^H
\mathbf A_i ) = \frac{P_{\rm t}}{M}, \, i = 1, \ldots, \frac{M}{2}.
\label{orgprobgeneralcase}
\end{eqnarray}
Introducing the new variables $ \mathbf X_k \triangleq \mathbf w_k \mathbf w_k^H,\ k=1,\ldots,K/2$
and following similar steps as in the case of two transmit waveforms,
the problem above can be equivalently rewritten as
\begin{eqnarray} \nonumber
&\min\limits_{\mathbf X_k} \max\limits_{ \theta_q}& \left| G_{\rm d}
(\theta_q)/2 - \sum_{k=1}^{K/2} {\rm tr} \bigg( \mathbf d (\theta_q
) \mathbf d^H (\theta_q) \mathbf X_k\bigg ) \right| \\
&{\rm s.t.}& \sum_{k=1}^{K/2} {\rm tr} ( \mathbf X_k
\mathbf A_i ) = \frac{P_{\rm t}}{M}, \, i = 1, \ldots, \frac{M}{2} \nonumber \\
&& {\rm rank}(\mathbf X_k)=1,\ \ k=1,\cdots,K/2
\label{eq:NonConvexOptgenearlcase}
\end{eqnarray}
where $\mathbf X_k, k=1,\cdots,K/2$ are Hermitian matrices. The
problem \eqref{eq:NonConvexOptgenearlcase} can be solved in a
similar way as the problem~\eqref{rank}. Specifically, the optimal
solution of the problem \eqref{eq:NonConvexOptgenearlcase} can be
approximated using the SDP relaxation, i.e., dropping the rank-one constraints
and solving the resulting convex problem.

By relaxing the rank-one constraints, the optimization problem
\eqref{eq:NonConvexOptgenearlcase} can be approximated as
\begin{eqnarray}\nonumber
&\min\limits_{\mathbf X_k} \max\limits_{ \theta_q}& \left| G_{\rm d}
(\theta_q)/2 - \sum_{k=1}^{K/2} {\rm tr} ( \mathbf d (\theta_q )
\mathbf d^H (\theta_q) \mathbf X_k) \right|  \\
&{\rm s.t.}& \sum_{k=1}^{K/2} {\rm tr} ( \mathbf X_k
\mathbf A_i ) = \frac{P_{\rm t}}{M}, \, i = 1, \ldots, \frac{M}{2} \nonumber \\
&& \mathbf X_k\succeq {\bf 0},\ \ k=1,\cdots,K/2.
\label{eq:ConvexOptgenearlcase}
\end{eqnarray}
The problem \eqref{eq:ConvexOptgenearlcase} is convex and,
therefore, it can be solved efficiently using interior point
methods. Once the matrices $\mathbf X_k\succeq {\bf 0},\ k=1,\cdots,K/2$
are obtained, the corresponding weight vectors $\mathbf w_k,\
k=1,\cdots,K/2$ can be obtained using randomization techniques.
Specifically, we use the randomization method introduced in
Subsection~\ref{BDes2Wave} over every $\mathbf X_k, k=1,\cdots,K/2$
separately and then map the resulted rank-one solutions to the
closest feasible points. Among the candidate solutions, the best one
is then selected.

\subsection{Optimal Rotation of the Transmit Beamspace Matrix}
The solution of the optimization problem
\eqref{orgprobgeneralcase} is not unique and as it will be explained
shortly in details, any spatial rotation of the optimal transmit
beamspace matrix is also optimal. Among the set of the optimal
solutions of the problem \eqref{orgprobgeneralcase}, the one with
better energy preservation is favorable. As a result, after the
approximate optimal solution of the problem
\eqref{orgprobgeneralcase} is obtained, we still need to find the
optimal rotation which results in the best possible transmit
beamspace matrix in terms of the energy preservation. More
specifically, since the DOA of the target at $\theta_l$ is estimated
based on the phase difference between the signal components of this
target in the newly defined vectors, i.e.,  $\sum_{i=1}^{K/2}
{\mathbf d}^H(\theta_l) {\mathbf w}_i$ and $\sum_{i=K/2+1}^{K} {\mathbf d}^H(\theta_l){\mathbf
w}_i$, to obtain the best performance, $\mathbf
W$ should be designed in a way that the magnitudes of the summations
$\sum_{i=1}^{K/2} {\mathbf d}^H(\theta_l)\mathbf w_i$ and
$\sum_{i=K/2+1}^{K} {\mathbf d}^H(\theta_l){\mathbf w}_i $ take their
largest values.

Since the phase of the product term ${\mathbf
d}^H(\theta_l) {\mathbf w}_i$ in $\sum_{i=1}^{K/2} {\mathbf d}^H(\theta_l){\mathbf w}_i$
(or equivalently in $\sum_{i=K/2+1}^{K} {\mathbf
d}^H(\theta_l) {\mathbf w}_i$) may be different for different waveforms, the terms in
the summation $\sum_{i=1}^{K/2} {\mathbf d}^H(\theta_l) {\mathbf w}_i$
(or equivalently in the summation $\sum_{i=K/2+1}^{K} {\mathbf d}^H(\theta_l){\mathbf w}_i$) may add incoherently and, therefore, it may result in
a small magnitude which in turn degrades the DOA estimation performance.
In order to avoid this problem, we use the property that any
arbitrary rotation of the transmit beamspace matrix does not change
the transmit beampattern. Specifically, if $ \mathbf W = [\mathbf
w_1,\cdots,\mathbf w_{K/2},\tilde{\mathbf
w}_1^*,\cdots,\tilde{\mathbf w}_{K/2}^*] $ is a transmit
beamspace matrix with the introduced structure, then the new
beamspace matrix defined as
\begin{equation}
\mathbf W_{\rm rot} = [\mathbf w_{\rm rot, 1},\cdots,\mathbf w_{\rm
rot, K/2},\tilde{\mathbf w}_{\rm rot, 1}^*,\cdots,\tilde{\mathbf
w}_{\rm rot, K/2}^*].
\end{equation}
has the same beampattern and the same power distribution across the antenna elements. Here $[\mathbf w_{\rm rot,
1},\cdots,\mathbf w_{\rm rot, K/2}]=[\mathbf w_1,\cdots,\mathbf
w_{K/2}] \mathbf U_{K/2 \times K/2}$ and $\mathbf U_{K/2 \times
K/2}$ is a unitary matrix. Based on this
property, after proper design of the beamspace matrix with a desired
beampattern and the RIP, we can rotate the beams so that the
magnitude of the summation $\sum_{i=1}^{K/2} {\mathbf
d}^H(\theta_l){\mathbf w}_i$ is increased as much as possible.

Since the actual locations of the targets are not known a priori, we
design a unitary rotation matrix so that the integration of the squared
magnitude of the summation $\sum_{i=1}^{K/2}  {\mathbf
d}^H(\theta_l){\mathbf w}_i$ over the desired sector is maximized. As an illustrating example and because of space limitations, we consider the case when $K$ is 4. In this case,
\begin{equation}
[\mathbf w_{\rm rot, 1},\mathbf w_{\rm rot, 2}]=[\mathbf w_1,\mathbf
w_{2}] \mathbf U_{2 \times 2} \label{unitary}
\end{equation}
and the integration of the squared magnitude of the summation
$\sum_{i=1}^{2}  {\mathbf d}^H(\theta_l){\mathbf w}_{\rm rot, i}$ over the desired
sectors can be expressed as
\begin{eqnarray}
\int_{\Theta}\!\!\!&\!\!\!&\!\!\!\!\!\!\!\bigg|\mathbf w_{\rm rot, 1}^H  \mathbf d(\theta)
\!\!+\!\! \mathbf w_{\rm rot, 2}^H  \mathbf d(\theta)  \bigg|^2 d \theta\nonumber\\
\!\!\!\!&\!\!\!=\!\!\!&\!\!\!\!\! \int_{\Theta}\!\!\bigg(\!\mathbf
d^H(\theta)\mathbf w_{\rm rot, 1} \mathbf w^H_{\rm rot, 1} \mathbf
d(\theta) \!+\! \mathbf d^H(\theta)\mathbf w_{\rm rot, 2}
\mathbf w^H_{\rm rot, 2} \mathbf d(\theta)  \nonumber \\
&&\ \ \ + 2 {\rm Re}\big(\mathbf d^H(\theta)\mathbf w_{\rm rot, 1}
\mathbf w^H_{\rm rot, 2} \mathbf d(\theta)\big)\!\bigg) d \theta
\nonumber
\\\!\!\!\!&\!\!\!=\!\!\!&\!\!\!\!\! \int_{\Theta}\!\bigg(\mathbf d^H(\theta)\mathbf w_{1}
\mathbf w^H_{1} \mathbf d(\theta) \!+\! \mathbf d^H(\theta)\mathbf
w_{2} \mathbf w^H_{2} \mathbf d(\theta)  \nonumber \\
&&\ \ + 2 {\rm Re}\big(\mathbf d^H(\theta)\mathbf w_{\rm rot, 1}
\mathbf w^H_{\rm rot, 2} \mathbf d(\theta)\big)\bigg) d \theta
\label{SSM}
\end{eqnarray}
where $\Theta$ denotes the desired sectors and ${\rm Re}(\cdot)$ stands
for the real part of a complex number. The last line follows from the
equation \eqref{unitary}. Defining the new vector $\mathbf e =[1,
-1]^T$, the expression in \eqref{SSM} can be equivalently recast as
\begin{eqnarray}
\int_{\Theta} \bigg( \!\!\!&\!\!\!\!\!&\!\!\! \mathbf d^H(\theta)\mathbf w_{1} \mathbf w^H_{1} \mathbf d(\theta) \!+\! \mathbf d^H(\theta)\mathbf w_{2} \mathbf w^H_{2} \mathbf d(\theta) \nonumber\\
\!\!&\!\!\!\!&\!\! + 2{\rm Re}\big(\mathbf d^H(\theta)\mathbf w_{\rm rot, 1} \mathbf w^H_{\rm rot, 2} \mathbf d(\theta)\big) \bigg) d \theta = \nonumber\\
  \int_{\Theta}  \bigg( \!\!\!&\!\!\!\!\!&\!\!\! 2\mathbf d^H(\theta)\mathbf w_{1}  \mathbf w^H_{1} \mathbf d(\theta) \!+\! 2 \mathbf d^H(\theta)\mathbf w_{2} \mathbf w^H_{2} \mathbf d(\theta) \nonumber\\
\!\!&\!\!\!\!&\!\!  -|\mathbf d(\theta)^H \mathbf W \mathbf U \mathbf
e|^2 \bigg) d \theta. \label{integsum}
\end{eqnarray}
We aim at maximizing the expression \eqref{integsum} with respect to
the unitary rotation matrix  $\mathbf U$. Since the first two terms
inside the integral in \eqref{integsum} are independent of the
unitary matrix, it only suffices to minimize the integration of the
last term.

Using the property that $\|\mathbf X\|_{\rm F}^2 = {\rm tr}(\mathbf
X \mathbf X^H)$, where $\| \cdot \|_{\rm F}$ denotes the Frobenius
norm, and the cyclical property of the trace, i.e., ${\rm tr}(\mathbf
X \mathbf X^H)={\rm tr}( \mathbf X^H \mathbf X)$, the integral of
the last term in \eqref{integsum} can be equivalently expressed as
\begin{equation}
\label{resultedintegGeneralCase} \int_\Theta {\rm tr}~\big(\mathbf U
\mathbf e \mathbf e^{H} \mathbf U^{H} \mathbf W^{H} \mathbf
d(\theta) \mathbf d(\theta)^{H} \mathbf W\big) \mathrm{d}\theta.
\end{equation}
The only term in the integral \eqref{resultedintegGeneralCase} which
depends on $\theta$ is $\mathbf W^H\mathbf d(\theta)\mathbf
d(\theta)^H\mathbf W$. Therefore, the minimization of the
integration of the last term in \eqref{integsum} over a sector
$\Theta$ can be stated as the following optimization problem
\begin{eqnarray}
\label{GrasGeneralCase}
\min_{\mathbf U} && {\rm tr}(\mathbf U \mathbf E \mathbf U^H \mathbf D) \\
{\rm s.t.} && \mathbf{U} \mathbf {U}^H = \mathbf{I} \nonumber
\end{eqnarray}
\\
where $\mathbf E= \mathbf e \mathbf e^{H}$ and $\mathbf D =
\int_\Theta {\rm tr}~\big( \mathbf W^{H} \mathbf d(\theta) \mathbf
d(\theta)^{H} \mathbf W\big) \mathrm{d}\theta$. Because of the
unitary constraint, the optimization problem \eqref{GrasGeneralCase}
is the optimization problem over the Grassmannian manifold
\cite{Grassmannian}, \cite{Grassmannian2}. In order to address this
problem, we can use the existing steepest descent-based
algorithm developed in \cite{Grassmannian}.

\subsection{Spatial-Division Based Design (SDD)}
It is worth noting that instead of designing all transmit beams jointly, an easy alternative for designing $\mathbf W$ is to design different
pairs of beamforming vectors $\{ \mathbf w_k,\ \tilde{\mathbf w}^{*}_k\}$,
$k=1,\cdots,K/2$ separately. Specifically, in order to avoid the
incoherent summation of the terms in $\sum_{i=1}^{K/2}  {\mathbf d}^H(\theta_l){\mathbf
w}_i$ (or equivalently in $\sum_{i=K/2+1}^{K}
{\mathbf d}^H(\theta_l){\mathbf w}_i $), the matrix $\mathbf W$ can be
designed in such a way that the corresponding transmit beampatterns
of the beamforming vectors $\mathbf w_1,\cdots,\mathbf w_{K/2}$ do
not overlap and they cover different parts of the desired sector
with equal energy. This alternative design is referred to as the SDD method.
The design of one pair $\{ \mathbf w_k,\ \tilde{\mathbf w}^{*}_k\}$ has been already
explained in Subsection~\ref{BDes2Wave}.

\section{Simulation Results}
Throughout our simulations, we assume a uniform linear transmit
array with $M=10$ antennas spaced half a wavelength apart, and a
non-uniform linear receive array of $N=10$ elements. The locations of the
receive antennas are randomly drawn from the set $[0,\ 9]$ measured
in half a wavelength. Noise signals are assumed to be Gaussian,
zero-mean, and white both temporally and spatially. In each example,
targets are assumed to lie within a given spatial sector. From
example to example the sector widths in which transmit energy is
focussed is changed, and, as a result, so does the optimal number of
waveforms to be used in the optimization of the transmit beamspace
matrix. The optimal number of waveforms is calculated based on the
number of dominant eigen-values of the positive definite matrix
$\mathbf A = \int_{\Theta}{\bf a}(\theta){\bf a}^H(\theta)d\theta$ (see \cite{TransmitEnergyFocusing} for explanations
and corresponding Cramer-Rao bound derivations and analysis).
We assume that the number of dominant eigenvalues is even;
otherwise, we round it up to the nearest even number. The reason that an odd number
of dominant eigenvalues is rounded up, as opposed to down, is that
overusing waveforms is less detrimental to the performance of DOA
estimation than underusing, as it is shown in
\cite{TransmitEnergyFocusing}. Four examples are chosen to test the
performance of our algorithm. In Example~1, a single centrally
located sector of width $20^ \circ$ is chosen to verify the
importance of the uniform power distribution across the orthogonal
waveforms. In Example~2, two separated sectors each with a width of
$20^ \circ$ degrees are chosen. In Example~3, a single, centrally
located sector of width $10^ \circ$ degrees is chosen. Finally, in
Example~4, a single, centrally located sector of width $30^ \circ$
degrees is chosen. The optimal number of waveforms used for each
example is two, four, two, and four, respectively. The methods
tested by the examples are traditional MIMO radar with uniform transmit
power density and $K = M$ and the proposed jointly optimum transmit beamspace
design method. In Example~3, we also consider the SSD method which is an easier alternative to the jointly optimal method.
Throughout the simulations, we refer to the proposed transmit
beamspace method as the optimal transmit beamspace design (although the solution obtained through SDP relaxation and randomization is suboptimal in general) to distinguish it from the SDD method in which different pairs of the transmit beamspace matrix columns are
designed separately. In Examples~1~and~3, the SDD is not considered since there
is no need for more than two waveforms. We also do not apply the SDD method in
the last example due to the fact that the corresponding spatially
divided sectors in this case are adjacent and their sidelobes
result in energy loss and performance degradation as opposed to Example~2.

Throughout all simulations, the total transmit power remains
constant at $P_{\rm t} = M$. The root mean square error (RMSE) and probability of target resolution are calculated based on $500$
independent Monte-Carlo runs.

\subsection{Example~1 : Effect of the Uniform Power Distribution Across the Waveforms}
\begin{figure}[h]
\centering \centerline{\epsfig{figure=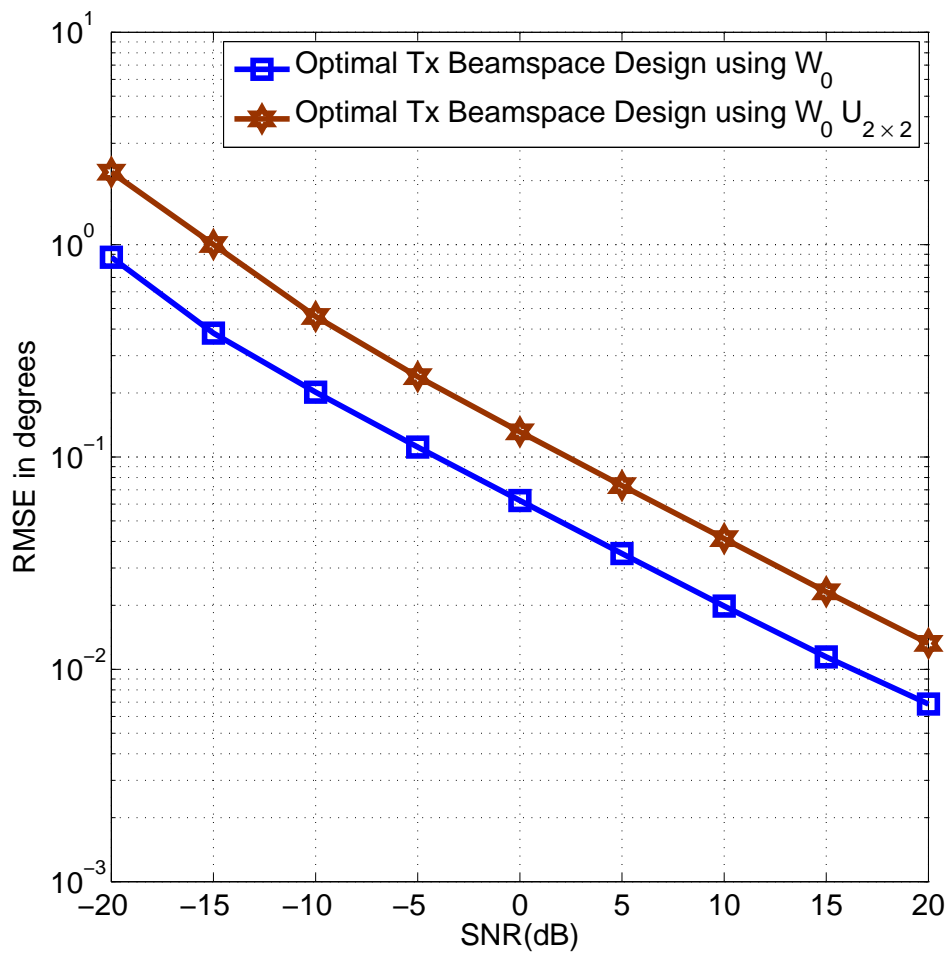,width=14.0cm}}
\caption{Example~1: Performance of the new proposed method with
and without uniform power distribution across transmit waveforms.}
\label{F22}
\end{figure}

In this example, we aim at studying how the lack of uniform
transmission power across the transmit waveforms affects the
performance of the new proposed method. For this goal, we consider
two targets that are located in the directions $-5^ \circ$ and $5^
\circ$ and the desired sector is chosen as $\theta = [-10^\circ\
10^\circ]$. Two orthogonal waveforms are considered and optimal
transmit beamspace matrix denoted as $\mathbf W_0$ is obtained by solving
the optimization problem \eqref{objective2}--\eqref{ME_PC}. To simulate the case of non-uniform power distribution across the waveforms while
preserving the same transmit beampattern of $\mathbf W_0$, we use the rotated transmit beamspace matrix
$\mathbf W_0 \mathbf U_{2 \times 2}$, where $\mathbf U_{2 \times 2}$
is a unitary matrix defined as
\begin{equation}
{\mathbf U}_{2 \times 2}= \begin{bmatrix}
 0.6925 + j 0.3994 & 0.4903 + j 0.3468 \\[0.3em]
 -0.4755 + j 0.3669 & 0.6753 - j 0.4279 \\[0.3em]
 \end{bmatrix}.\nonumber
\end{equation}
Note that $\mathbf W_0$ and $\mathbf W_0 {\mathbf U}_{2 \times 2}$ lead to the same
transmit beampattern and as a result the same transmit power within the
desired sector, however, compared to the former, the latter one does
not have uniform transmit power across the waveforms. The
RMSE curves of the proposed DOA estimation method for both $\mathbf W_0$ and $\mathbf W_0 {\mathbf U}_{2 \times 2}$ versus
${\rm SNR}$ are shown in Fig.~\ref{F22}. It can be seen from this
figure that the lack of uniform transmission power across the
waveforms can degrade the performance of DOA estimation severely.

\subsection{Example~2 : Two Separated Sectors of Width $20^\circ$ Degrees Each}
\begin{figure}[t]
\centerline{\includegraphics[width=14.0cm]{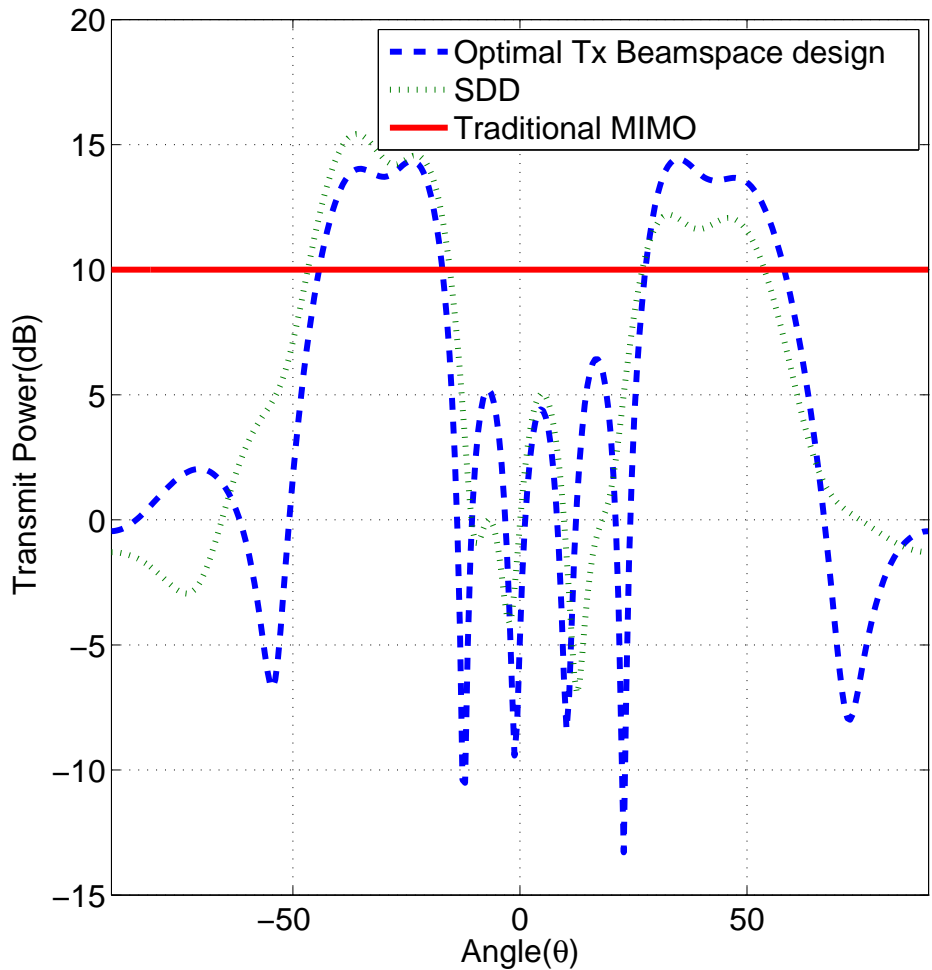}}
\caption{Example~2: Transmit beampatterns of the traditional MIMO
and the proposed transmit beamspace design-based methods.}
\label{fi:Ex1TransmitBeampattern}
\end{figure}

\begin{figure}[t!]
\centerline{\includegraphics[width=14.0cm]{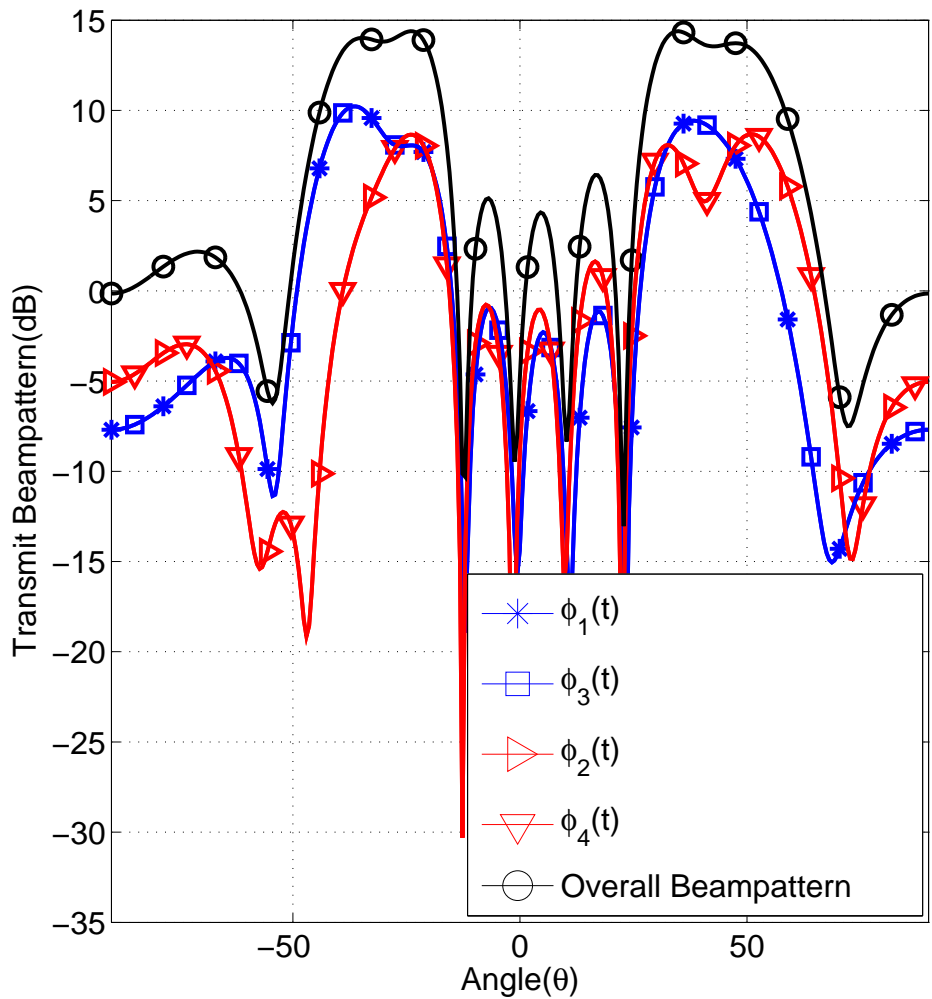}}
\caption{Example~2: Individual beampatterns associated with
individual waveforms and the overall beampattern.}
\label{fi:Ex1CohAddition}
\end{figure}

In the second example, two targets are assumed to lie within two
spatial sectors: one from $\theta = [-40^\circ\ -20^\circ]$ and the
other from $\theta = [30^\circ\ 50^\circ]$. The targets are located at $\theta_1 = -33^\circ$ and $\theta_2 = 41^\circ$.
Fig.~\ref{fi:Ex1TransmitBeampattern} shows the transmit beampatterns
of the traditional MIMO with uniform transmit power distribution
and both the optimal and SDD designs for ${\bf W}$. It can be seen in the figure
that the optimal transmit beamspace method provides the most even
concentration of power in the desired sectors. The SDD technique
provides concentration of power in the desired sectors above and
beyond traditional MIMO; however, the energy is not evenly
distributed with one sector having a peak beampattern strength of
15~dB, while the other has a peak of no more than 12~dB.
Fig.~\ref{fi:Ex1CohAddition} shows the individual beampatterns
associated with individual waveforms as well as the coherent
addition of all four individual beampatterns.

\begin{figure}[h!]
\centerline{\includegraphics[width=14.0cm]{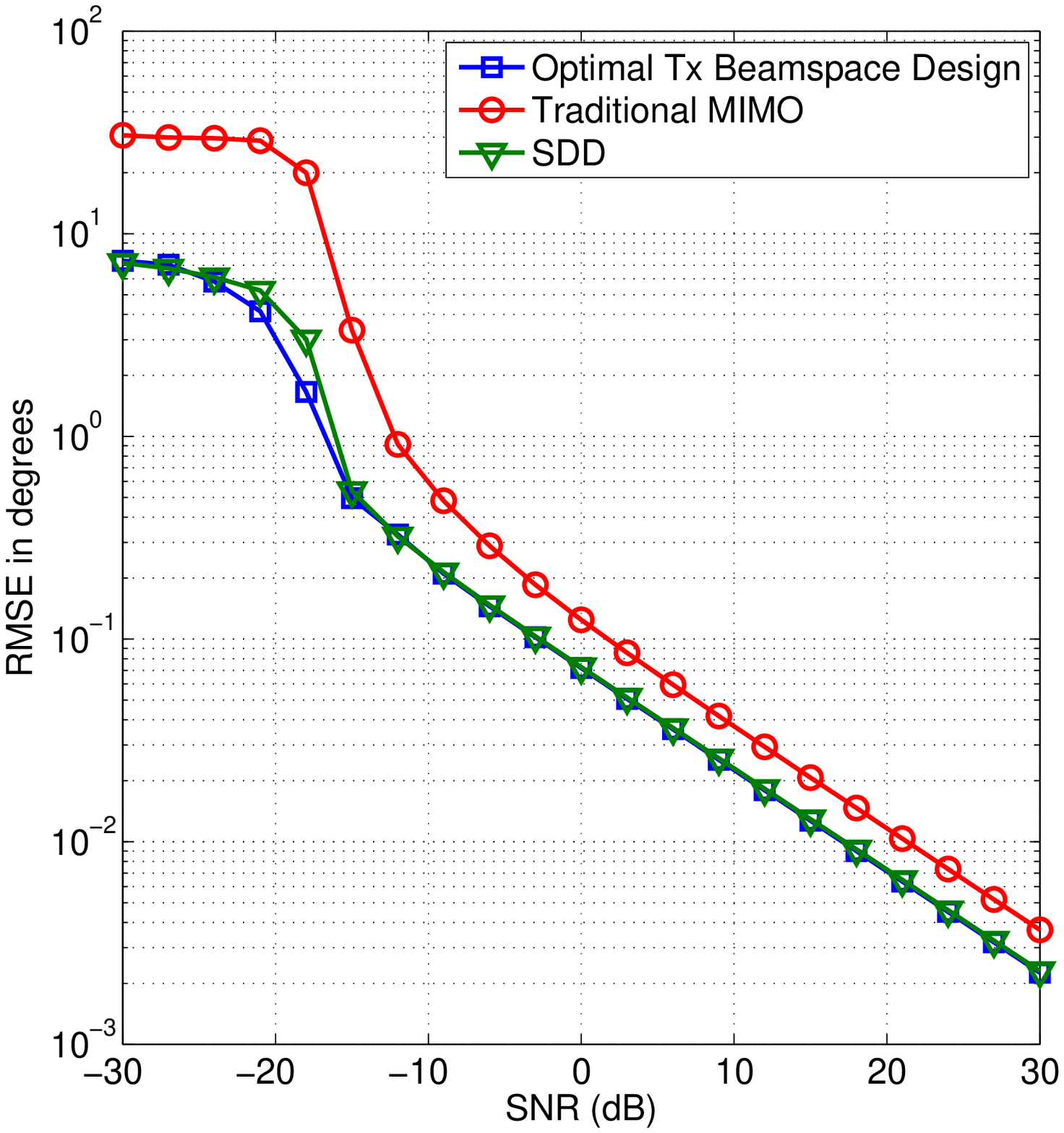}}
\caption{Example~2: Performance comparison between the traditional
MIMO and the proposed transmit beamspace design-based methods.}
\label{fi:Ex1RMSE}
\end{figure}

The performance of all three methods is compared in terms of the
corresponding RMSEs versus SNR as shown
in Fig.~\ref{fi:Ex1RMSE}. As we can see in the figure, the jointly optimal transmit beamspace and
the SDD methods have lower RMSEs as compared to the RMSE of the traditional
MIMO radar. It is also observed from the figure that
the performance of the SDD method is very close to the
performance of the jointly optimal one.

\begin{figure}[h!]
\centerline{\includegraphics[width=14.0cm]{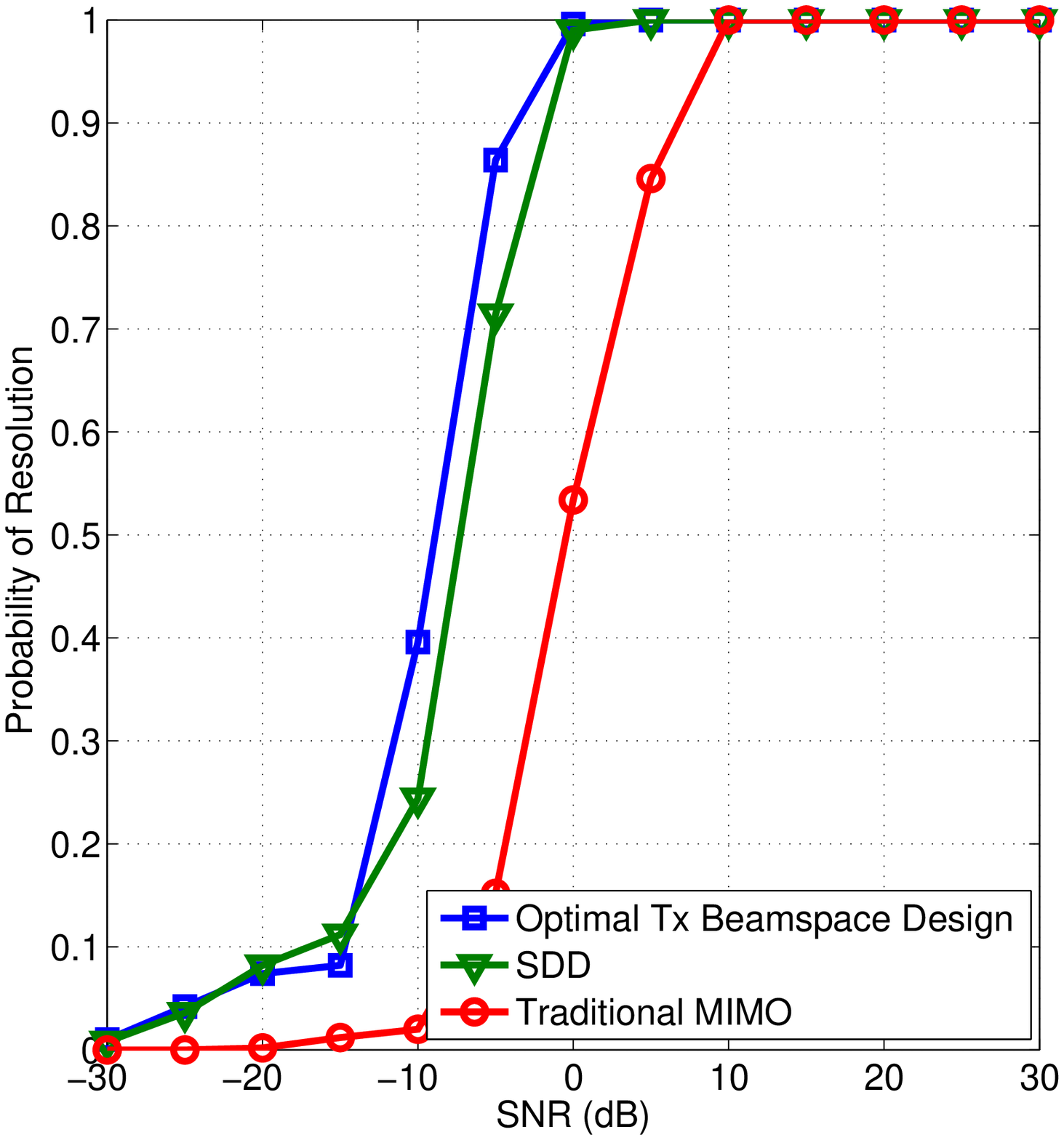}}
\caption{Example~2: Performance comparison between the traditional
MIMO and the proposed transmit beamspace design-based
methods.}\label{fi:Ex1Resolution}
\end{figure}

To assess the proposed method's ability to resolve closely located
targets, we move both targets to the locations $\theta_1 = 38^\circ$
and $\theta_2 = 40^\circ$.  The performance of all three methods
tested is given in terms of the probability of target
resolution. Note that the targets are considered to be resolved if
there are at least two peaks in the MUSIC spectrum and the following
is satisfied \cite{VanTrees}
\begin{equation}\label{eq:resolution}
\left|\hat\theta_l - \theta_l\right|\leq \frac{\Delta\theta}{2},
\quad l=1,2\nonumber
\end{equation}
where $\Delta\theta = |\theta_2-\theta_1|$. The probability of
source resolution versus SNR for all methods tested are shown in
Fig.~\ref{fi:Ex1Resolution}. It can be seen from the figure that
the SNR threshold at which the probability of target resolution
transitions from very low values (i.e., resolution fail) to values
close to one (i.e., resolution success)  is lowest for the jointly optimal
transmit beamspace design-based method, second lowest for the
SDD method, and finally, highest for the traditional
MIMO radar method. In other words, the figure shows that the jointly optimal transmit
beamspace design-based method has a higher probability of target
resolution at lower values of SNR than the SDD method, while the traditional MIMO radar method has the worst
resolution performance.


\subsection{Example~3 : Single and Centrally Located Sector of Width $10^\circ$ Degrees}
\begin{figure}[t!]
\centerline{\includegraphics[width=14.0cm]{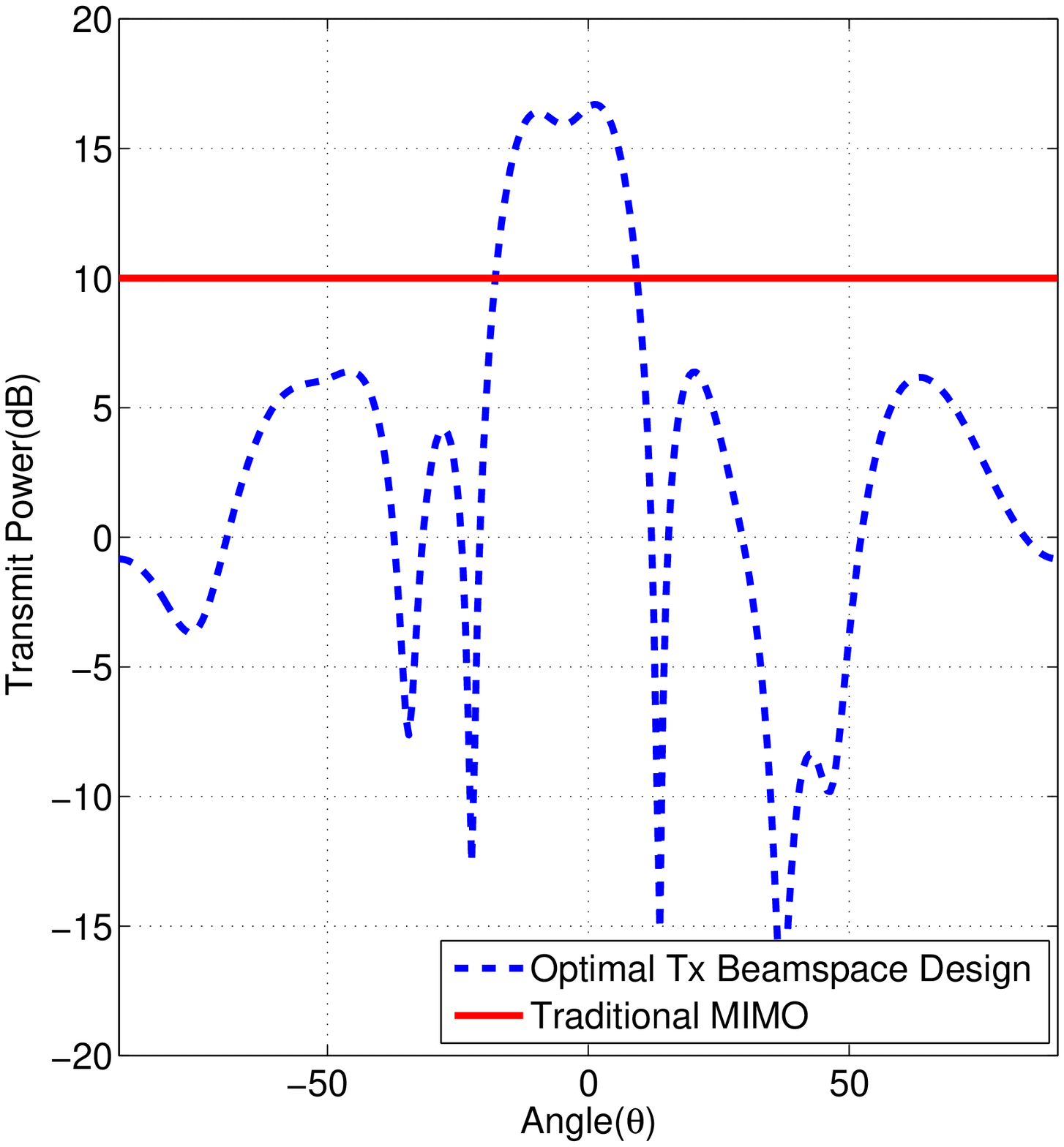}}
\caption{Example~3: Transmit beampatterns of the traditional MIMO
and the proposed transmit beamspace design-based
method.}\label{fi:ThinSectorBeampattern}
\end{figure}

\begin{figure}[h]
\centerline{\includegraphics[width=14.0cm]{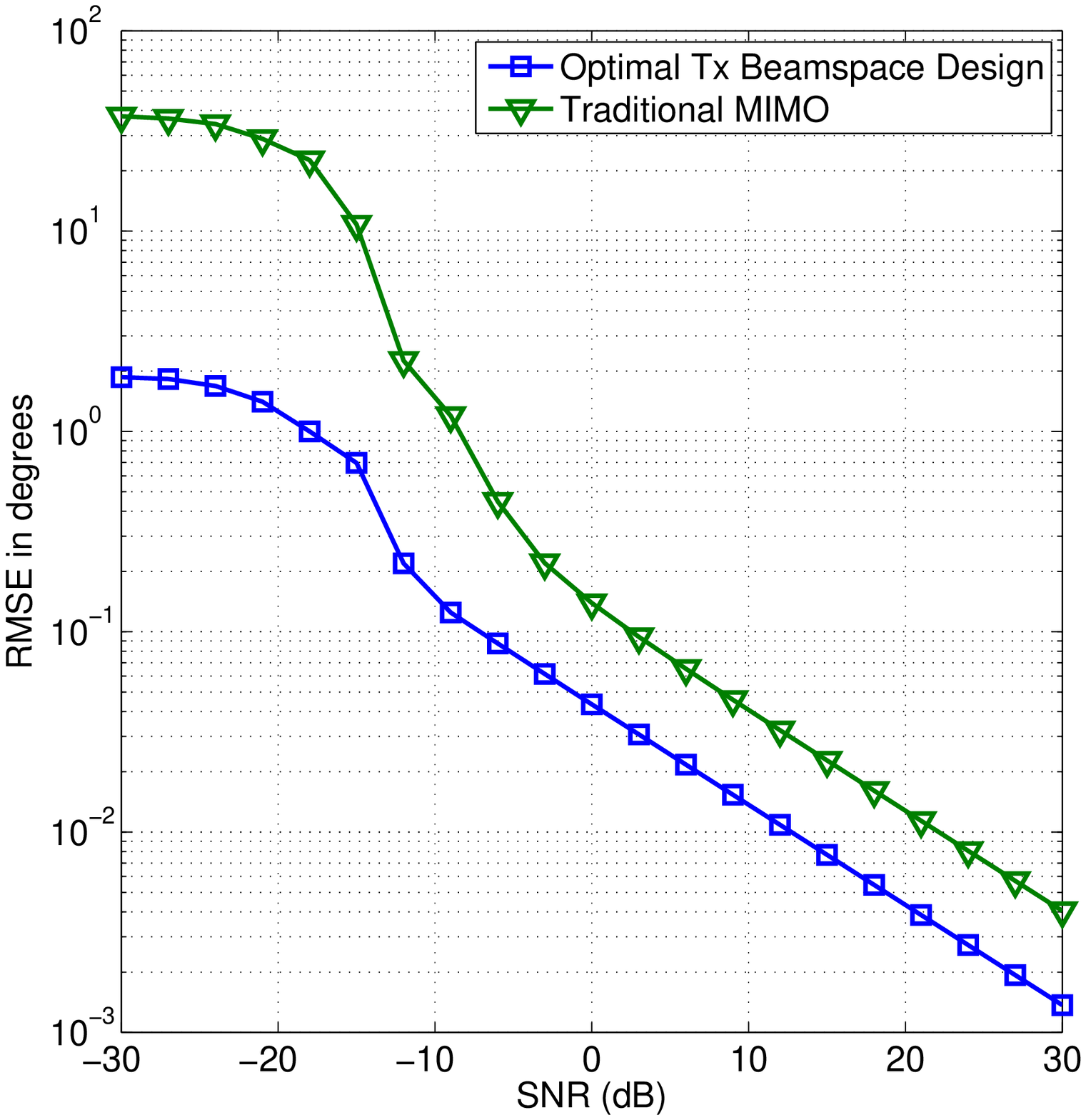}}
\caption{Example~3: Performance comparison between the traditional
MIMO and the proposed transmit beamspace design-based
method.}\label{fi:RMSEvsSNRexample2}
\end{figure}

\begin{figure}[h!]
\centerline{\includegraphics[width=14.0cm]{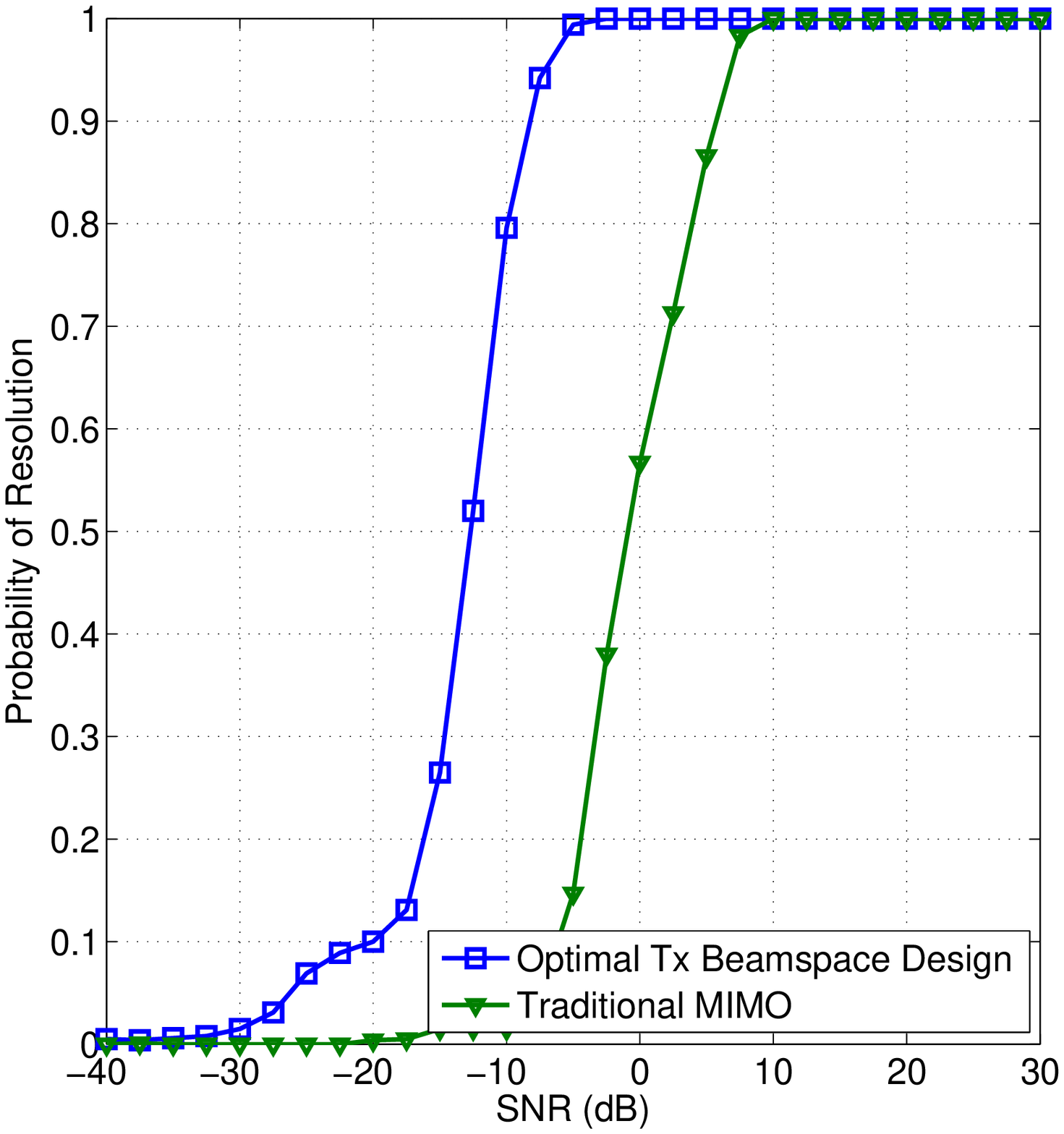}}
\caption{Example~3: Performance comparison between the traditional
MIMO and the proposed transmit beamspace design-based
methods.}\label{fi:Ex2Resolution}
\end{figure}

In the third example, the targets are assumed to lie within a
single thin sector of $\theta = [-10^\circ\ 0^\circ]$. Due to the
choice of the width of the sector, the optimal number of waveforms to
use is only two. For this reason, only two methods are tested: the
proposed transmit beamspace method and the traditional MIMO radar. The
beampatterns for these two methods are shown in
Fig.~\ref{fi:ThinSectorBeampattern}. It can observed from the figure
that our method offers a transmit power gain that is 5~dB higher
than the traditional MIMO radar.  In order to test the RMSE
performance of both methods, targets are assumed to be located at $\theta_1 =
-7^\circ$ and $\theta_2 = -2^\circ$. The RMSE's are plotted versus
SNR in Fig.~\ref{fi:RMSEvsSNRexample2}. It can be observed from this
figure that the proposed transmit beamspace method yields lower RMSE as compared to the traditional MIMO radar based method
at moderate and high SNR values. At low SNR values one can observe
from the figure that the RMSE of the transmit beamspace method
saturates at $3^{\circ}$ due to the fact that each of the two targets
is located $3^{\circ}$ from the edge of the sector. In order to test
the resolution capabilities of both methods tested, the targets are moved to $\theta_1 = -3^\circ$ and
$\theta_2 = -1^\circ$. The same criterion as in Example~2 is then used
to determine the target resolution. The results of this test are
displayed in Fig.~\ref{fi:Ex2Resolution} and agrees with the similar results in Example~2.


\subsection{Example~4 : Single and Centrally Located Sector of Width $30^\circ$ Degrees}

\begin{figure}[t!]
\centerline{\includegraphics[width=14.0cm]{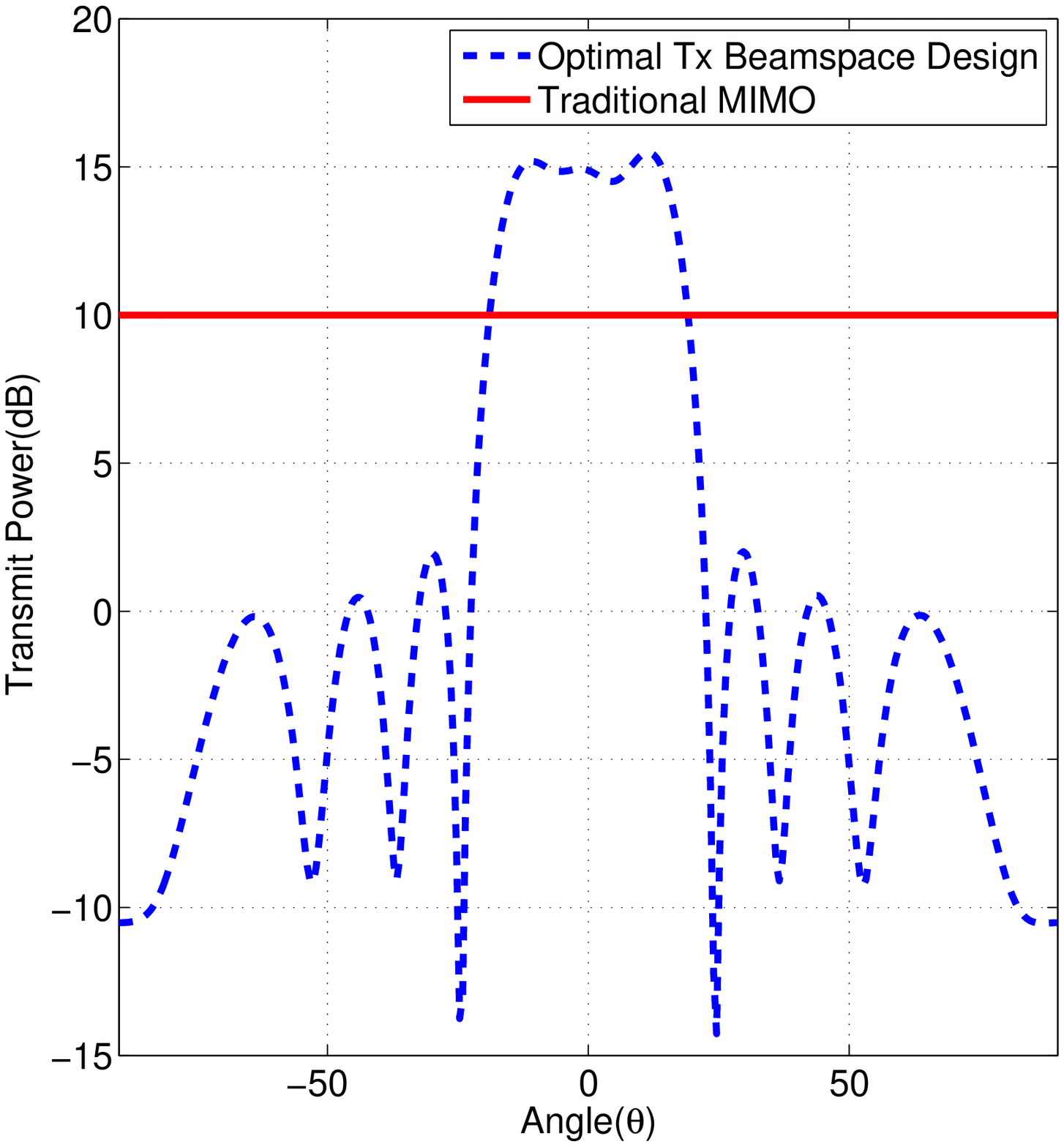}}
\caption{Example~4: Transmit beampatterns of the traditional MIMO
and the proposed methods.} \label{fi:Ex3Beampattern}
\end{figure}

\begin{figure}[t!]
\centerline{\includegraphics[width=14.0cm]{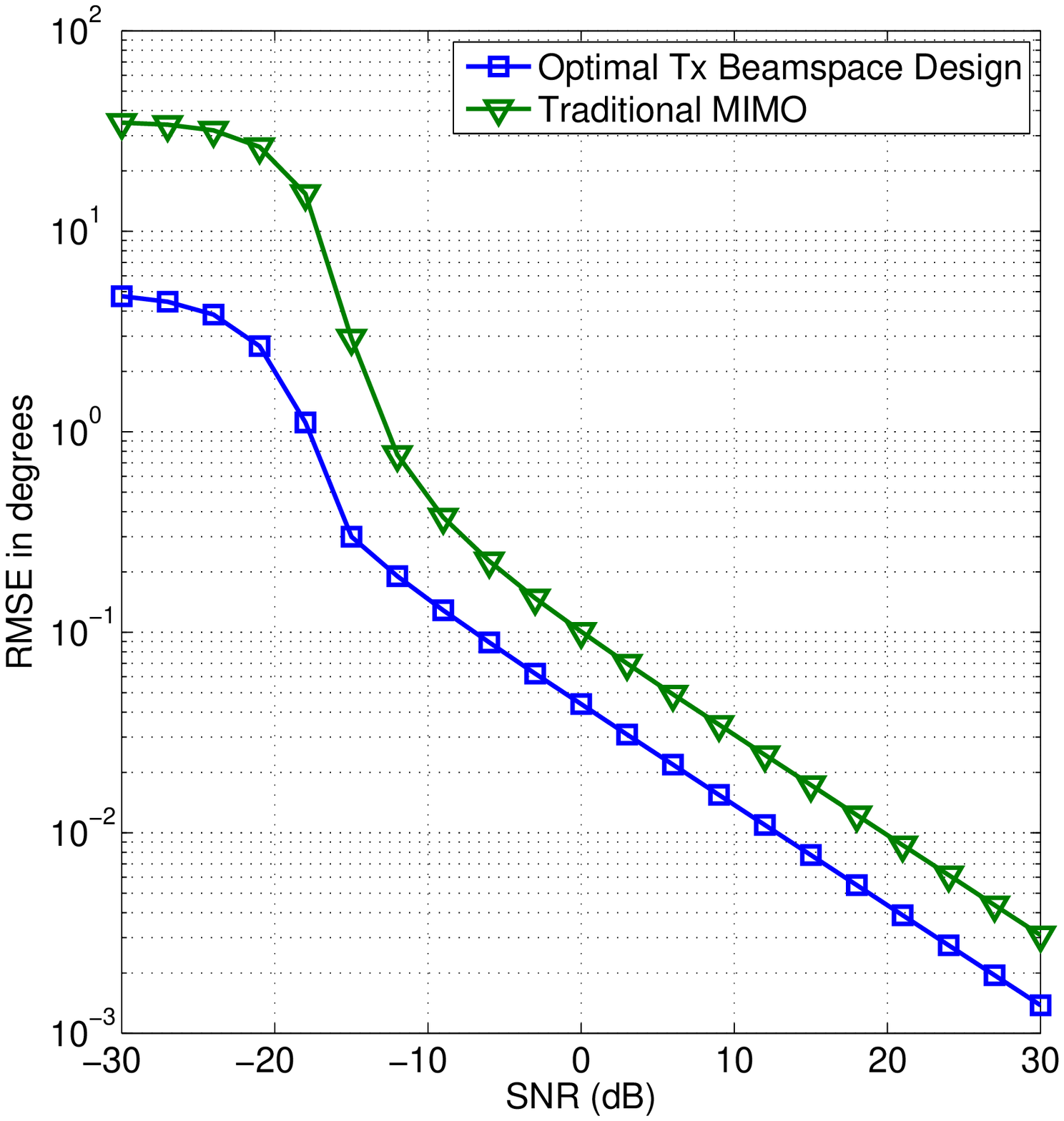}}
\caption{Example~4: Performance comparison between the traditional
MIMO and the proposed methods.} \label{fi:Ex3RMSEvsSNR}
\end{figure}

In the last example, a single wide sector is chosen as $\theta =
[-15^\circ\ 15^\circ]$. The optimal number of waveforms for such a
sector is found to be four. Similar to the previous Example~3, we
compare the performance of the proposed method to that of the
traditional MIMO radar. Four transmit beams are used to simulate the optimal transmit beamspace design-based method.
Fig.~\ref{fi:Ex3Beampattern} shows the transmit beampatterns for the
methods tested.
In order to test the RMSE performance of the methods tested, two
targets are assumed to be located at $\theta_1 = -12^\circ$ and
$\theta_2 = 9^\circ$. Fig.~\ref{fi:Ex3RMSEvsSNR} shows the RMSEs versus SNR for the methods tested. As we can see
in the figure, the RMSE for the jointly optimal transmit beamspace design-based
method is lower than the RMSE for the traditional MIMO radar based method.
Moreover, in order to test resolution, the targets are moved to
$\theta_1 = -3^\circ$ and $\theta_2 = -1^\circ$. The same criterion
as in Example~2 is used to determine the target resolution. The results
of this test are similar to those displayed in
Fig.~\ref{fi:Ex1Resolution}, and, therefore, are not displayed here.


\section{Conclusion}
The problem of transmit beamspace design for MIMO radar with colocated antennas with
application to DOA estimation has been considered. A new method for designing the transmit beamspace
matrix that enables the use of search-free DOA estimation techniques
at the receiver has been introduced. The essence of the proposed
method is to design the transmit beamspace matrix based on
minimizing the difference between a desired transmit beampattern and
the actual one. The case of even but otherwise arbitrary number of
transmit waveforms has been considered. The transmit beams are designed in
pairs where all pairs are designed jointly while satisfying the requirements that the two transmit beams
associated with each pair enjoy rotational invariance with respect
to each other. Unlike previous methods that achieve phase rotation
between two transmit beams while allowing the magnitude to be
different, a specific beamspace matrix structure achieves phase rotation while ensuring
that the magnitude response of the two transmit beams is exactly the same at all spatial directions has been proposed.
The SDP relaxation technique has been used to transform the proposed formulation into
a convex optimization problem that can be solved efficiently  using interior
point methods. An alternative SDD method
that divides the spatial domain into several subsectors and
assigns a subset of the transmit beamspace pairs to each subsector
has been also developed. The SDD method enables post processing of data
associated with different subsectors independently with DOA
estimation performance comparable to the performance of the joint
transmit beamspace design-based method. Simulation results have been used to
demonstrate the improvement in the DOA estimation performance
offered by using the proposed joint and SDD transmit beamspace
design methods as compared to the traditional MIMO radar.


\begin{thebibliography}{10}
\bibitem{Viberg}
H.~Krim and M.~Viberg, ``Two decades of array signal processing
research: the parametric approach,'' {\it IEEE Signal Processing
Mag.}, vol.~13, no.~4, pp.~67--94, Aug.~1996.
%
\bibitem{VanTrees}
H.~Van~Trees, {\it Optimum Array Processing.} Willey, 2002.
%
\bibitem{An_idea} E.~Fishler, A.~Haimovich, R.~Blum, D.~Chizhik, L.~Cimini,
and  R.~ Valenzuela, ``{MIMO} radar: An idea whose time has come,'' in
\emph{Proc. IEEE Radar Conf.}, Honolulu, Hawaii, USA, Apr. 2004,
vol.~2, pp. 71--78.
%
\bibitem{Jli}
J.~Li and P.~Stoica, \textit{MIMO Radar Signal Processing}. New
Jersy: Wiley, 2009.
%
\bibitem{Radar_separated_review} A.~Haimovich, R.~Blum, and L.~Cimini, ``{MIMO} radar
with widely separated antennas,'' \emph{{IEEE} Signal Processing
Mag.}, vol.~25, pp. 116--129, Jan. 2008.
%
\bibitem{DeMaio2008} A.~De~Maio, M.~Lops, and L.~Venturino, ``Diversity-integration tradeoffs in MIMO detection,'' {\it IEEE Trans. Signal
Processing}, vol.~56, no.~10, pp.~5051--5061, Oct.~2008.
%
\bibitem{NasrSergGershman} A.~Hassanien, S.~A.~Vorobyov, and
A.~B.~Gershman, ``Moving target parameters estimation in
non-coherent MIMO radar systems,'' {\it IEEE Trans. Signal
Processing}, vol.~60, no. 5, pp.~2354--2361, May~2012.
%
\bibitem{MIMORadarSensitivity_Nehorai} M.~Akcakaya and A.~Nehorai, ``MIMO radar sensitivity
analysis for target detection,''{\it IEEE Trans. Signal Processing}, vol.~59, no.~7, pp.~3241--3250, Jul.~2011.
%
\bibitem{Radar_colocated_review} J.~Li and P.~Stoica, ``{MIMO} radar
with colocated antennas,'' \emph{{IEEE} Signal Processing Mag.},
vol.~24, pp. 106--114, Sept. 2007.
%
\bibitem{NasrSergConf} A.~Hassanien and S.~A.~Vorobyov,
``Transmit/receive beamforming for MIMO radar with colocated
antennas,'' in {\it Proc. IEEE Inter. Conf. Acoustics, Speech, and
Signal Processing}, Taipei, Taiwan, Apr.~2009, pp.~2089-2092.
%
\bibitem{Vaidyanathan_Pal} P.~P.~Vaidyanathan and P.~Pal, ``MIMO radar, SIMO radar,
and IFIR radar: A comparison,'' in {\it Proc. 63rd Asilomar Conf. Signals, Syst. and Comput.},
 Pacific Grove, CA, Nov.~2009, pp.~160--167.
%
\bibitem{PhasedMIMOradar} A.~Hassanien and S.~A.~Vorobyov, ``Phased-MIMO radar:
A tradeoff between phased-array and MIMO radars,'' {\it IEEE
Trans. Signal Processing}, vol.~58, no.~6, pp.~3137--3151,
Jun.~2010.
%
\bibitem{EUSIPCO} A.~Hassanien and S.~A.~Vorobyov, ``Why the
phased-MIMO radar outperforms the phased-array and MIMO radars,''
in {\it Proc. 18th European Signal Processing Conf.}, Aalborg,
Denmark, Aug.~2010, pp.~1234--1238.
%
\bibitem{SubarrayedTxBeamformingUK} D.~Wilcox and M.~Sellathurai, ``On MIMO radar
subarrayed transmit beamforming," \emph{IEEE Trans. Signal Processing}, vol.~60, no.~4, pp.~2076--2081, Apr.~2012.
%
\bibitem{TransmitEnergyFocusing} A.~Hassanien and S.~A.~Vorobyov,, ``Transmit energy
focusing for DOA estimation in MIMO radar with colocated
antennas,'' {\it IEEE Trans. Signal Processing}, vol.~59, no.~6,
pp.~2669--2682, Jun.~2011.
%
\bibitem{Abeysekera2013} G.~Hua and S.~S.~Abeysekera, ``Receiver design for range and doppler sidelobe suppression using MIMO and phased-array radar,'' {\it IEEE Trans. Signal Processing}, vol.~61, no.~6, pp.~1315--1326, Mar.~2013.
%
\bibitem{Duofang}
C.~Duofang, C.~Baixiao, and Q.~Guodong, ``Angle estimation using
ESPRIT in MIMO radar,'' {\it Electron. Lett.}, vol.~44, no.~12,
pp.~770--771, Jun.~2008.

\bibitem{NionNikos}
D.~Nion and N.~D.~Sidiropoulos, ``Tensor algebra and multidimensional harmonic retrieval in signal processing for MIMO
radar,'' {\it IEEE Trans. Signal Processing}, vol.~58, no.~11,
pp.~5693--5705, Nov.~2010.
%
\bibitem{HybridMIMOphased} D.~Fuhrmann, J.~Browning, M.~Rangaswamy, ``Signaling strategies
for the hybrid {MIMO} phased-array radar," \emph{IEEE J. Sel. Topics Signal Processing}, vol.~4, no.~1, pp.~66-–78, Feb.~2010.
%
\bibitem{Fuhrmann}
D.~Fuhrmann and G.~San Antonio, ``Transmit beamforming for MIMO
radar systems using signal cross-correlation,'' {\it IEEE Trans.
Aerospace and Electronic Systems}, vol.~44, no.~1, pp.~171--186,
Jan.~2008.
%
\bibitem{koivunen}
T.~Aittomaki and V.~Koivunen, ``Beampattern optimization by minimization
of quartic polynomial,'' in Proc. {\it 15 IEEE/SP Statist. Signal
Process. Workshop}, Cardiff, U.K., Sep. 2009, pp. 437–440.
%
\bibitem{Stoica} H.~He, P.~Stoica,  and J.~Li, ``Designing unimodular sequence sets with good correlations--Including an application to MIMO radar,'' \emph{IEEE Trans. Signal Processing}, vol.~57, no.~11,  pp.4391--4405, Nov.~2009.
%
\bibitem{NasrSergCAMSAP} A.~Hassanien and S.~A.~Vorobyov, ``Direction finding
for MIMO radar with colocated antennas using transmit beamspace
preprocessing,'' in {\it Proc. IEEE Int. Workshop
on Computational Advances in Multi-Sensor Adaptive Processing (CAMSAP'09)}, Aruba, Dutch
Antilles, Dec.~2009, pp.~181--184.
%
\bibitem{Tarokh}
V.~Tarokh, H.~Jafarkhani, and A.~R.~Calderbank, ``Space-time block
codes from orthogonal designs,'' {\it IEEE Trans. Inf. Theory},
vol.~45, no.~7, pp.~1456Ð1467, Jul.~1999.
%
\bibitem{ArashSergiy}
A.~Khabbazibasmenj, S.~A.~Vorobyov, and A.~Hassanien, ``Transmit
beamspace design for direction finding in colocated MIMO radar
with arbitrary receive array," in {\it Proc. 36th IEEE Inter.
Conf. Acoustics, Speech, and Signal Processing}, Prague, Czech
Republic, May~2011, pp. 2784-2787.
%
\bibitem{SergiyArashMatt}
A.~Khabbazibasmenj, S.~A.~Vorobyov, A.~Hassanien, M.W.~Morency,
``Transmit beamspace design for direction finding in colocated
MIMO radar with arbitrary receive array and even number of
waveforms,'' in {\it Proc. 46th Asilomar Conf. Signals, Syst, and
Comput.}, Pacific Grove, CA, Nov.~4-7, 2012.
%
\bibitem{SDP2}
Z.-Q.~Luo, W.-K.~Ma,  A.~M.-C.~So, Y.~Ye, and S.~Zhang,
``Semidefinite Relaxation of Quadratic Optimization Problems,''
{\it IEEE Signal Processing Mag.}, vol.~27, no.~3, pp.~20-34,
May~2010.
\bibitem{Boyd}
A.~d'Aspremont and S.~Boyd, ``Relaxation and randomized method for
nonconvex QCQPs,'' class note,  {\texttt http://www.stanford.edu/class/ee392o/}.
%
\bibitem{SDR}
H.~Wolkowicz, ``Relaxations of Q2P,'' in {\it Handbook of
Semidefinite Programming: Theory, Algorithms, and Applications},
H. Wolkowicz, R. Saigal, and L.Venberghe, Eds. Norwell, MA:
Kluwer, 2000, ch. 13.4.
%
\bibitem{ArashSergiy3}
 A.~Khabbazibasmenj, S.~A.~Vorobyov, and A.~Hassanien, ``Robust adaptive beamforming based on steering vector estimation with as little as possible prior information,'' {\it IEEE Trans. Signal Processing}, vol.~60, no.~6, pp.~2974--2987, Jun.~2012.
 %
%
\bibitem{SDP3}
K.~T.~Phan, S.~A.~Vorobyov, N.~D.~Sidiropoulos, and C.~Tellambura,
``Spectrum sharing in wireless networks via QoS-aware secondary
multicast beamforming,'' {\it IEEE Trans. Signal Processing},
vol.~57, no.~6, pp.~2323-2335, Jun.~2009.
%
\bibitem{Grassmannian}
T.~E.~Abrudan, J.~Eriksson, and V.~Koivunen, ``Steepest descent
algorithms for optimization under unitary matrix constraint,'' {\it
IEEE Trans. Signal Process.}, vol.~56, no.~3, pp.~1134-–1147, Mar.~
2008.

\bibitem{Grassmannian2}
P.~-A.~Absil, R.~Mahony, and R.~Sepulchre, ``Riemannian geometry of
Grassmann manifolds with a view on algorithmic computation,'' {\it
Acta Applicandae Mathematicae}, vol.~80, no.~2, pp.~199–-220, 2004.


\end{thebibliography}
\end{document}